\documentclass[10pt,a4paper,onecolumn]{article}

\usepackage{algorithm2e}
\usepackage{amsmath}
\usepackage{amssymb}
\usepackage{authblk}
\usepackage[english]{babel}
\usepackage{changes}
\usepackage{color}
\usepackage[plain]{fancyref}
\usepackage[margin=3cm]{geometry}
\usepackage{graphicx}
\usepackage[colorlinks]{hyperref}
\usepackage{microtype}
\usepackage{natbib}
\usepackage{todonotes}
\usepackage{verbatim}

\newcommand{\ourtitle}{Evolving to learn:\\discovering interpretable plasticity rules for spiking networks}
\newcommand{\ourkeywords}{metalearning, learning to learn, synaptic plasticity, spiking neuronal networks, genetic programming}

\definecolor{darkblue}{rgb}{0.0,0.0,0.4}
\definecolor{darkgreen}{rgb}{0.0,0.4,0.0}
\definecolor{darkred}{rgb}{0.6,0.0,0.0}
\definecolor{parametergray}{gray}{0.8}
\hypersetup{colorlinks, 
  linkcolor=darkred,
  citecolor=darkred,
  urlcolor=darkred,
  pdftitle=\ourtitle{},
  pdfauthor={Jakob Jordan, Maximilian Schmidt, Walter Senn, Mihai A.~Petrovici},
  pdfkeywords=\ourkeywords{},
}

\newcommand*{\fancyrefalglabelprefix}{alg}
\frefformat{plain}{\fancyrefalglabelprefix}{algorithm~#1}
\Frefformat{plain}{\fancyrefalglabelprefix}{Algorithm~#1}
\frefformat{vario}{\fancyrefalglabelprefix}{algorithm~#1}
\Frefformat{vario}{\fancyrefalglabelprefix}{Algorithm~#1}
\frefformat{margin}{\fancyrefalglabelprefix}{alg.~#1}
\Frefformat{margin}{\fancyrefalglabelprefix}{Alg.~#1}
\makeatletter
\def\mkfancyprefix#1#2{%
  \@namedef{fancyref#1labelprefix}{#1}%
  \begingroup\def\x{\endgroup\frefformat{plain}}%
  \expandafter\x\csname fancyref#1labelprefix\endcsname
  {\MakeLowercase{#2}\fancyrefdefaultspacing##1}%
  \begingroup\def\x{\endgroup\Frefformat{plain}}%
  \expandafter\x\csname fancyref#1labelprefix\endcsname
  {#2\fancyrefdefaultspacing##1}%
}
\makeatother

\mkfancyprefix{sub}{Sec.}
\mkfancyprefix{sec}{Sec.}
\mkfancyprefix{fig}{Fig.}
\mkfancyprefix{alg}{Alg.}
\mkfancyprefix{eq}{Eq.\!}

\definecolor{JJ}{RGB}{250,50,50}
\definecolor{MS}{RGB}{21,176,26}
\definechangesauthor[color=JJ]{JJ}
\definechangesauthor[color=MS]{MS}
\definechangesauthor[color=orange]{MAP}
\definechangesauthor[color=red]{WS}

\newcommand{\e}{\mathrm{e}}

\newcommand{\ms}{\,\mathrm{ms}}
\newcommand{\Tinter}{T_{\mathrm{inter}}}
\newcommand{\Tpattern}{T_{\mathrm{pattern}}}
\newcommand{\SNR}{\mathrm{SNR}}
\newcommand{\whom}{w^{\mathrm{hom}}}

\newcommand{\Ejr}{E_j^\text{r}}
\newcommand{\Ravg}{\bar{R}}
\newcommand{\Ravgplus}{\bar{R}^+}
\newcommand{\Ravgminus}{\bar{R}^-}
\newcommand{\Ravgabs}{\bar{R}_\text{abs}}

\newlength{\columnwidthleft}
\newlength{\columnwidthmiddle}
\newlength{\columnwidthmiddleconn}

\usepackage{tabularx}       % professional-quality tables
\usepackage{xcolor}

\newcommand{\Hz}{\,\text{Hz}}

\newcommand{\mV}{\,\text{mV}}

\newcommand{\pA}{\,\text{pA}}

\newcommand{\pF}{\,\text{pF}}

\newcommand{\psp}{\bar s}

\definecolor{white}{RGB}{255,255,255}
\definecolor{white}{RGB}{100,100,100}

\newcommand{\modelhdr}[3]{
  \multicolumn{#1}{|l|}{
    \textbf{\makebox[0pt][l]{#2}\hspace{0.5\textwidth}\makebox[0pt][c]{#3}}
  }
}

\def\Cplusplus{{C\nolinebreak[4]\hspace{-.05em}\raisebox{.4ex}{\tiny\bf ++}}\hspace{.2em}}

\title{\ourtitle}

\date{\vspace{-1em}}
\author[1]{Jakob Jordan\thanks{These authors contributed equally to this work.}}
\author[2,3]{Maximilian Schmidt$^*$}
\author[1]{Walter Senn}
\author[1,4]{Mihai A.~Petrovici}
\affil[1]{Department of Physiology, University of Bern, Bern, Switzerland}
\affil[2]{Ascent Robotics, Tokyo, Japan}
\affil[3]{RIKEN Center for Brain Science, Tokyo, Japan} 
\affil[4]{Kirchhoff-Institute for Physics, Heidelberg University, Heidelberg, Germany}

\begin{document}

\maketitle

\begin{abstract} % <=150 words for eLife
  Continuous adaptation allows survival in an ever-changing world.
  Adjustments in the synaptic coupling strength between neurons are essential for this capability, setting us apart from simpler, hard-wired organisms.
  How these changes can be mathematically described at the phenomenological level, as so called ``plasticity rules'', is essential both for understanding biological information processing and for developing cognitively performant artificial systems.
  We suggest an automated approach for discovering biophysically plausible plasticity rules based on the definition of task families, associated performance measures and biophysical constraints.
  By evolving compact symbolic expressions we ensure the discovered plasticity rules are amenable to intuitive understanding, fundamental for successful communication and human-guided generalization.
  We successfully apply our approach to typical learning scenarios and discover previously unknown mechanisms for learning efficiently from rewards, recover efficient gradient-descent methods for learning from target signals, and uncover various functionally equivalent STDP-like rules with tuned homeostatic mechanisms.
\end{abstract}


\section{Introduction}

\begin{figure}[h]
  \centering
  \includegraphics[width=1.\textwidth]{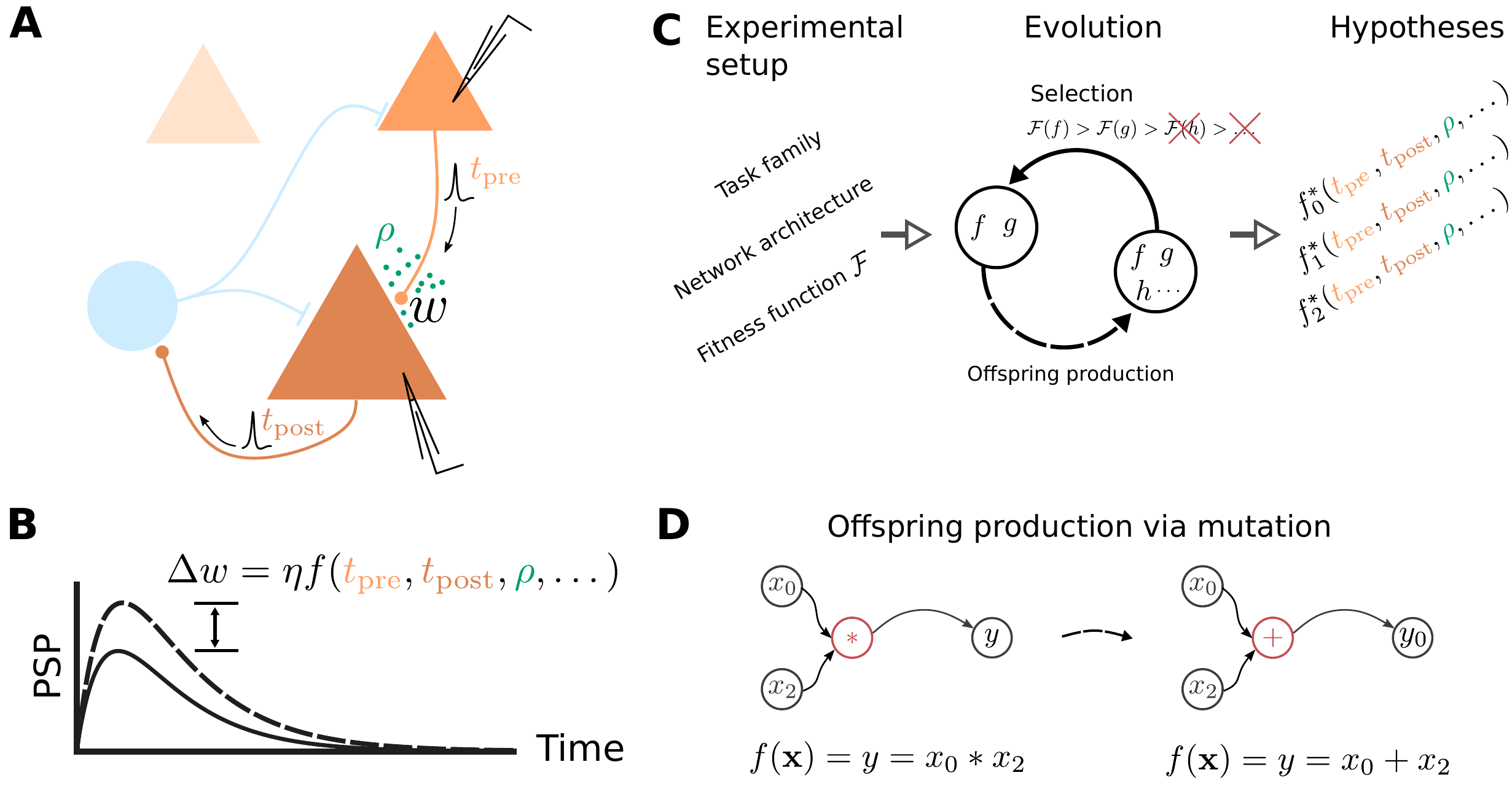}
  \caption{{\bf Artificial evolution of synaptic plasticity rules in spiking neuronal networks.}
    {\bf (A)} Sketch of cortical microcircuits consisting of pyramidal cells (orange) and inhibitory interneurons (blue).
    Stimulation elicits action potentials in pre- and postsynaptic cells, which, in turn, influence synaptic plasticity.
    {\bf (B)} Synaptic plasticity leads to a weight change ($\Delta w$) between the two cells, here measured by the change in the amplitude of post-synaptic potentials.
    The change in synaptic weight can be expressed by a function $f$ that in addition to spike timings ($t_\text{pre}, t_\text{post}$) can take into account additional local quantities, such as the concentration of neuromodulators ($\rho$, green dots in A) or postsynaptic membrane potentials.
    {\bf (C)} For a specific experimental setup, an evolutionary algorithm searches for individuals representing functions $f$ that maximize the corresponding fitness function $\mathcal{F}$.
    An offspring is generated by modifying the genome of a parent individual.
    Several runs of the evolutionary algorithm can discover phenomenologically different solutions ($f_0,f_1,f_2$) with comparable fitness.
    {\bf (D)} An offspring is generated from a single parent via mutation. Mutations of the genome can, for example, exchange mathematical operators, resulting in a different function $f$.
  }\label{fig:intro-microcircuit}
\end{figure}

How do we learn?
Whether we are memorizing the way to the lecture hall at a conference or mastering a new sport, somehow our central nervous system is able to retain the relevant information over extended periods of time, sometimes with ease, other times only after intense practice.
This acquisition of new memories and skills manifests at various levels of the system, with changes of the interaction strength between neurons being a key ingredient.
Uncovering the mechanisms behind this synaptic plasticity is a key challenge in understanding brain function.
Most studies approach this monumental task by searching for phenomenological models described by symbolic expressions that map local biophysical quantities to changes of the connection strength between cells (\Fref{fig:intro-microcircuit}A, B).

Approaches to deciphering synaptic plasticity can be broadly categorized into bottom-up and top-down.
Bottom-up approaches typically rely on experimental data \citep[e.g.,][]{artola1990different,dudek1993bidirectional,bi1998synaptic,ngezahayo2000synaptic} to derive dynamic equations for synaptic parameters that lead to functional emergent macroscopic behavior if appropriately embedded in networks \citep[e.g.,][]{gutig2003learning,izhikevich2007solving,clopath2010connectivity}.
Top-down approaches proceed in the opposite direction: from a high-level description of network function, e.g., in terms of an objective function \citep[e.g.,][]{toyoizumi2005generalized,deneve2008bayesianinference,kappel2015network,kutschireiter2017nonlinear,sacramento2018dendritic,goltz2019fast}, dynamic equations for synaptic changes are derived and biophysically plausible implementations suggested.
Evidently, this demarcation is not strict, as most approaches seek some balance between experimental evidence, functional considerations and model complexity.
However, the relative weighting of each of these aspects is usually not made explicit in the communication of scientific results, making it difficult to track by other researchers.
Furthermore, the selection of specific tasks to illustrate the effect of a suggested learning rule is usually made only after the rule was derived based on other considerations.
Hence, this typically does not consider competing alternative solutions, as an exhaustive comparison would require significant additional investment of human resources.
A related problem is that researchers, in a reasonable effort to use resources efficiently, tend to focus on promising parts of the search space around known solutions, leaving large parts of the search space unexplored \citep{radi2003discovering}.
Automated procedures, in contrast, can perform a significantly less biased search.

We suggest an automated approach to discover learning rules in spiking neuronal networks that explicitly addresses these issues.
Automated procedures interpret the search for biological plasticity mechanisms as an optimization problem \citep{bengio1992optimization}, an idea typically referred to as meta-learning or learning-to-learn.
These approaches make the emphasis of particular aspects that guide this search explicit and place the researcher at the very end of the process, supporting much larger search spaces and the generation of a diverse set of hypotheses.
Furthermore, they have the potential to discover domain-specific solutions that are more efficient than general-purpose algorithms.
Early experiments focusing on learning in artificial neural networks (ANNs) made use of gradient descent or genetic algorithms to optimize parameterized learning rules \citep{bengio1990learning,bengio1992optimization,bengio1993generalization} or genetic programming to evolve less constrained learning rules \citep{bengio1994use,radi2003discovering}, rediscovering mechanisms resembling the backpropagation of errors \citep{linnainmaa1970representation,ivakhnenko1971polynomial,rumelhart1985learning}.
Recent experiments demonstrate how optimization methods can design optimization algorithms for recurrent ANNs \citep{andrychowicz2016learning}, evolve machine learning algorithms from scratch \citep{real2020automl}, and optimize parametrized learning rules in neuronal networks to achieve a desired function \citep{confavreux2020meta}.

We extend these meta-learning ideas to discover free-form, yet interpretable plasticity rules for spiking neuronal networks.
The discrete nature of spike-based neuronal interactions endows these networks with rich dynamical and functional properties \citep[e.g.,][]{dold2019stochasticity,jordan2019deterministic,keup2020transient}.
In addition, with the advent of non-von Neumann computing systems based on spiking neuronal networks with online learning capabilities \citep{moradi2017scalable,davies2018loihi,billaudelle2019versatile}, efficient learning algorithms for spiking systems become increasingly relevant for non-conventional computing.
Here, we employ genetic programming \citep[\Fref{fig:intro-microcircuit}C, D;][]{koza2010human} as a search algorithm for two main reasons.
First, genetic programming can operate on analytically tractable mathematical expressions describing synaptic weight changes that are interpretable.
Second, an evolutionary search does not need to compute gradients in the search space, thereby circumventing the need to estimate a gradient in non-differentiable systems.

We successfully apply our approach, which we refer to as ``evolving-to-learn'' (E2L), to three different learning paradigms for spiking neuronal networks: reward-driven, error-driven and correlation-driven learning.
For the reward-driven task our approach discovers new plasticity rules with efficient reward baselines perform that perform competively and even outperform previously suggested methods.
The analytic form of the resulting expressions suggests experimental approaches that would allow us to distinguish between them.
In the error-driven learning scenario, the evolutionary search discovers a variety of solutions which, with appropriate analysis of the corresponding expressions, can be shown to effectively implement stochastic gradient descent.
Finally, in the correlation-driven task, our method generates a variety of STDP kernels and associated homeostatic mechanisms that lead to similar network-level behavior.
This sheds new light onto the observed variability of synaptic plasticity and thus suggests a reevaluation of the reported variety in experimentally-measured STDP curves with respect to their possible functional equivalence.

Our results demonstrate the significant potential of automated procedures in the search for plasticity rules in spiking neuronal networks, analogous to the transition from hand-designed to learned features that lies at the heart of modern machine learning.

\section{Results}

\subsection{Setting up an evolutionary search for plasticity rules}

We introduce the following recipe to search for biophysically plausible plasticity rules in spiking neuronal networks.
First we determine a task family of interest and an associated experimental setup which includes specification of the network architecture, e.g., neuron types and connectivity, as well as stimulation protocols or training data sets.
Crucially, this step involves defining a fitness function to guide the evolutionary search towards promising regions of the search space.
It assigns high fitness to those individuals, i.e., learning rules, that solve the task well and low fitness to others.
The fitness function may additionally contain constraints implied by experimental data or arising from computational considerations.
We determine each individual's fitness on various examples from the given task family, e.g., different input spike train realizations, to discover plasticity rules that generalize well \citep{chalmers1991evolution, soltoggio2018born}.
Finally, we specify the neuronal variables available to the plasticity rule, such as low-pass-filtered traces of pre- and postsynaptic spiking activity or neuromodulator concentrations.
This choice is guided by biophysical considerations, e.g., which quantities are locally available at a synapse, as well as by the task family, e.g., whether reward or error signals are provided by the environment.
We write the plasticity rule in the general form $\Delta w = \eta \, f(\dots)$, where $\eta$ is a fixed learning rate, and employ an evolutionary search to discover functions $f$ that lead to high fitness.

\begin{figure}[tbp]
  \centering
  \includegraphics[width=1.\textwidth]{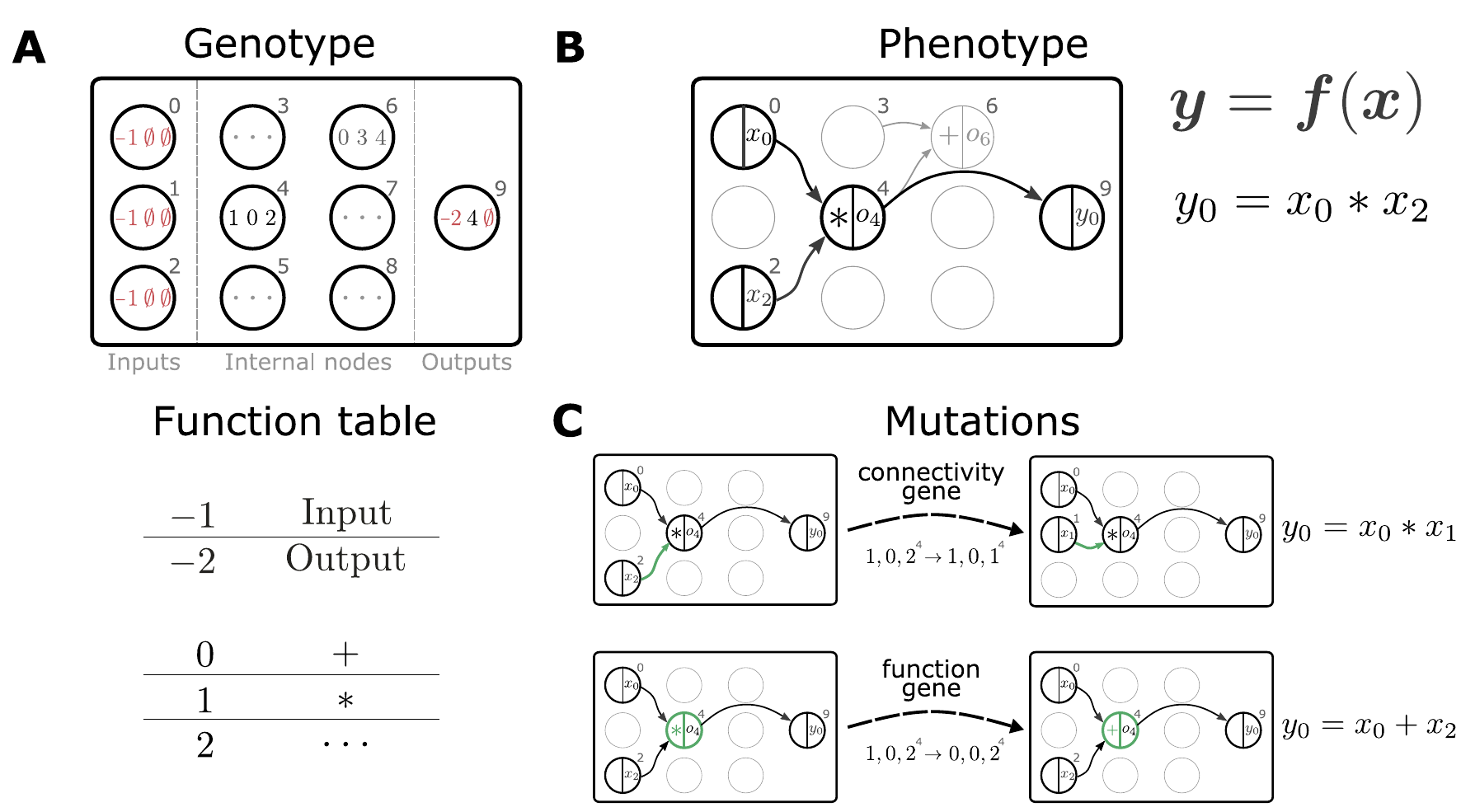}
  \caption{{\bf Representation and mutation of mathematical expressions in Cartesian genetic programming.}
    {\bf (A)} The genotype of an individual is a two-dimensional Cartesian graph (top).
    In this example, the graph contains three input nodes ($0-2$), six internal nodes ($3-8$) and a single output node ($9$).
    In each node the genes of a specific genotype are shown, encoding the operator used to compute the node's output and its inputs.
    Each operator gene maps to a specific mathematical function (bottom).
    Special values ($-1, -2$) represent input and output nodes.
    For example, node $4$ uses the operator $1$, the multiplication operation ``$*$'', and receives input from nodes $0$ and $2$.
    This node's output is hence given by $x_0 * x_2$.
    The number of input genes per node is determined by the operator with the maximal arity (here two).
    Fixed genes that cannot be mutated are highlighted in red.
    $\emptyset$ denotes non-coding genes.
    {\bf (B)} The computational graph (phenotype) generated by the genotype in A.
    Input nodes (${x_0, x_1, x_2}$) represent the arguments of the function $f$.
    Each output node selects one of the other nodes as a return value of the computational graph, thus defining a function from input $\boldsymbol{x}$ to output $\boldsymbol{y} = \boldsymbol{f}(\boldsymbol{x})$.
    Here, the output node selects node $4$ as a return value.
    Some nodes defined in the genotype are not used by a particular realization of the computational graph (in light gray, e.g., node $6$). Mutations that affect such nodes have no effect on the phenotype and are therefore considered ``silent''.
    {\bf (C)} Mutations in the genome either lead to a change in graph connectivity (top, green arrow) or alter the operators used by an internal node (bottom, green node).
    Here, both mutations affect the phenotype and are hence not silent.
  }\label{fig:cgp-sketch}
\end{figure}

We propose to use genetic programming (GP) as an evolutionary algorithm to discover plasticity rules in spiking neuronal networks.
GP applies mutations and selection pressure to an initially random population of computer programs to artificially evolve algorithms with desired behaviors \citep[e.g.,][]{koza1992genetic,koza2010human}.
Here we consider the evolution of mathematical expressions.
We employ a specific form of GP, Cartesian genetic programming \citep[CGP; e.g,][]{miller2000cartesian,miller2011cartesian}, that uses an indexed graph representation of programs.
The genotype of an individual is a two-dimensional Cartesian graph (\Fref{fig:cgp-sketch}A, top).
Over the course of an evolutionary run, this graph has a fixed number of input, output, and internal nodes.
The operation of each internal node is fully described by a single function gene and a fixed number of input genes.
A function table maps function genes to mathematical operations (\Fref{fig:cgp-sketch}A, bottom), while input genes determine from where this node receives data.
A given genotype is decoded into a corresponding computational graph (the phenotype, \Fref{fig:cgp-sketch}B) which defines a function $f$.
During the evolutionary run, mutations of the genotype alter connectivity and node operations, which can modify the encoded function (\Fref{fig:cgp-sketch}C).
The indirect encoding of the computational graph via the genotype supports variable-length phenotypes, since some internal nodes may not be used to compute the output (\Fref{fig:cgp-sketch}B).
The size of the genotype, in contrast, is fixed, thereby restricting the maximal size of the computational graph and ensuring compact, interpretable mathematical expressions.
Furthermore, the separation into genotype and phenotype allows the buildup of ``silent mutations'', i.e., mutations in the genotype that do not alter the phenotype.
These silent mutations can lead to more efficient search as they can accumulate and in combination lead to an increase in fitness once affecting the phenotype \citep{miller2000cartesian}.
A $\mu + \lambda$ evolution strategy \citep{beyer2002evolution} drives evolution by creating the next generation of individuals from the current one via tournament selection, mutation and selection of the fittest individuals (\Fref{sec:methods-evolutionary-algorithm}).
Prior to starting the search, we choose the mathematical operations that can appear in the plasticity rule and other (hyper)parameters of the Cartesian graph and evolutionary algorithm.
For simplicity, we consider a restricted set of mathematical operations and additionally make use of nodes with constant output.
After the search has completed, e.g., by reaching a target fitness value or a maximal number of generations, we analyze the discovered set of solutions.

In the following, we describe the results of three experiments following the recipe outlined above.

\subsection{Evolving an efficient reward-driven plasticity rule}
\label{sec:evolving-optimal-reward-driven-plasticity}

\begin{figure}
  % Skript: ../code/reward_learning/create_*.py
  % Hashes: 6f5a460844, b7ab3994d0, 4931d81e37, a0dc2d54ba
  \centering
  \includegraphics[width=1.\textwidth]{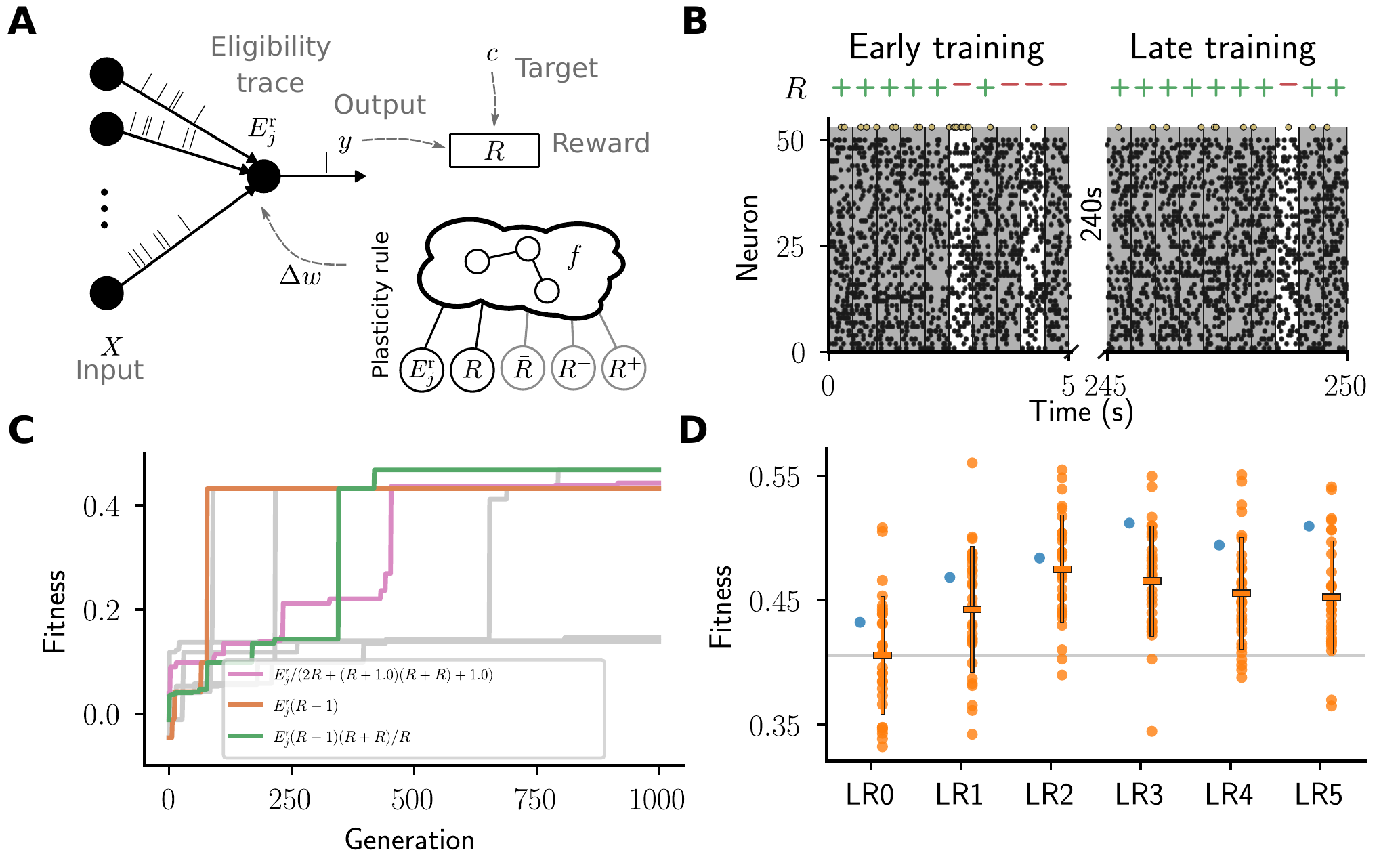}
  \caption{{\bf Cartesian genetic programming evolves various efficient reward-driven learning rules.}
    {\bf (A)} Network sketch.
    Multiple input neurons with Poisson activity project to a single output unit.
    Pre- and postsynaptic activity generate an eligibility trace in each synapse.
    Comparison between the output activity and the target activity generates a reward signal.
    $\Ravg$, and $\Ravgplus$, $\Ravgminus$ represent the expected reward, the expected positive and the expected negative reward, respectively.
    Depending on the hyperparameter settings either the former or the latter two are provided to the plasticity rule.
    {\bf (B)} Raster plot of the activity of input neurons (small black dots) and output neuron (large golden dots).
    Gray (white) background indicate patterns for which the output should be active (inactive).
    Top indicates correct classifications ($+$) and incorrect classifications ($-$).
    We show $10$ trials at the beginning (left) and the end of training (right) using the evolved plasticity rule: $\Delta w_j = \eta \, (R-1)\Ejr$.
    {\bf (C)} Fitness of best individual per generation as a function of the generation index for multiple example runs of the evolutionary algorithm with different initial conditions but identical hyperparameters.
    Labels show the expression $f$ at the end of the respective run for three runs resulting in well-performing plasticity rules.
    Gray lines represent runs with functionally identical solutions or low final fitness.
    {\bf (D)} Fitness of a selected subset of evolved learning rules on the $10$ experiments used during the evolutionary search (blue) and additional $30$ fitness evaluations, each on $10$ new experiments consisting of sets of frozen noise patterns and associated class labels not used during the evolutionary search (orange).
    Horizontal boxes represent mean, error bars indicate one standard deviation over fitness values.
    Gray line indicates mean fitness of LR$0$ for visual reference.
    See main text for the full expressions for all learning rules.
  }\label{fig:results-reinforcement-learning-fitness}
\end{figure}

We consider a simple reinforcement learning task for spiking neurons.
The experiment can be succinctly described as follows: $N$ inputs project to a single readout modeled by a leaky integrator neuron with exponential postsynaptic currents and stochastic spike generation (for details see \Fref{sec:methods-reinforcement-learning-task}).
We generate a finite number $M$ of frozen-Poisson-noise patterns of duration $T$ and assign each of these randomly to one of two classes.
The output neuron should learn to classify each of these spatio-temporal input patterns into the corresponding class using a spike/no-spike code (\Fref{fig:results-reinforcement-learning-fitness}A, B).

The fitness $\mathcal{F}(f)$ of an individual encoding the function $f$ is measured by the mean reward per trial averaged over a certain number of experiments $n_\text{exp}$, each consisting of $n$ classification trials
\begin{align}
  \label{eq:reward-learning-fitness}
  \mathcal{F}(f) := \frac{1}{n_\text{exp}} \sum_{k=1}^{n_\text{exp}} R_k(f) \; ,
\end{align}
where $R_k(f):=\frac{1}{n} \sum_{i=1}^{n} R_{k,i}(f)$ is the mean reward per trial obtained in experiment $k$.
The reward $R_{k,i}$ is the reward obtained in the $i$th trial of experiment $k$.
It is $1$ if the classification is correct and $-1$ otherwise.
In the following we drop the subscripts from $R_{k,i}$ where its meaning is clear from context.
Each experiment contains different realizations of frozen-noise input spike trains with associated class labels.

Previous work on reward-driven learning \citep{williams1992simple} has demonstrated that in policy-gradient-based approaches \citep[e.g.,][]{sutton2018reinforcement}, subtracting a so called ``reward baseline'' from the received reward can improve the convergence properties by adjusting the magnitude of weight updates.
However, choosing a good reward baseline is notoriously difficult \citep{williams1988toward,dayan1991reinforcement,weaver2001optimal}.
For example, in a model for reinforcement learning in spiking neurons, \citet{vasilaki2009spike} suggest the expected positive reward as a suitable baseline.
Here, we consider plasticity rules which, besides immediate rewards, have access to expected rewards.
These expectations are obtained as moving averages over a number of consecutive trials during one experiment and can either be estimated jointly ($\Ravg \in [-1, 1]$) or separately for positive ($\Ravgplus \in [0, 1]$) and negative ($\Ravgminus \in [-1, 0]$) rewards, with $\Ravg = \Ravgplus + \Ravgminus$ (for details, see \Fref{sec:methods-reinforcement-learning-task}).
In the former case, the plasticity rule takes the general form
\begin{align}
  \label{eq:reward-learning-gp-rule}
  \Delta w_j = \eta \, f \left( R, \Ejr(T), \Ravg \right) \; ,
\end{align}
while for seperately estimated positive and negative rewards it takes the form
\begin{align}
  \label{eq:reward-learning-gp-rule-2}
  \Delta w_j = \eta \, f \left( R, \Ejr(T), \Ravgplus, \Ravgminus \right) \; .
\end{align}
In both cases, $\eta$ is a fixed learning rate and $\Ejr(t)$ is an eligibility trace that contains contributions from the spiking activity of pre- and post-synaptic neurons which is updated over the course of a single trial (for details see \Fref{sec:methods-reinforcement-learning-task}).
The eligibility trace arises as a natural consequence of policy-gradient methods aiming to maximize the expected reward \citep{williams1992simple} and is a common ingredient of reward-modulated plasticity rules for spiking neurons \citep{vasilaki2009spike,fremaux2016neuromodulated}.
It is a low-pass filter of the product of two terms: the first is positive if the neuron was more active than expected by synaptic input; this can happen because the neuronal output is stochastic (\Fref{sec:methods-reinforcement-learning-task}), to promote exploration.
The second is a low-pass filter of presynaptic activity.
A simple plasticity rule derived from maximizing the expected rewards would, for example, change weights according to the product of the received reward and the eligibility trace: $\Delta w_j = R \Ejr$.
If by chance a neuron is more active than expected, and the agent receives a reward, all weights of active afferents are increased, making it more likely for the neuron to fire in the future given identical input.

Reward and eligibility trace values at the end of each trial ($t=T$) are used to determine synaptic weight changes.
In the following we drop the time argument of $\Ejr$ for simplicity.
Using CGP with three ($R$, $\Ejr, \Ravg$), or four inputs ($R$, $\Ejr, \Ravgplus, \Ravgminus$), respectively, we search for plasticity rules that maximize the fitness $\mathcal{F}(f)$ (\Fref{eq:reward-learning-fitness}).

None of the evolutionary runs with access to the expected reward ($\Ravg$) make use of it as a reward baseline (see Appendix \Fref{sec:app-reward-learning} for full data).
Some of them discover high-performing rules identical to that suggested by \citet{urbanczik2009reinforcement}: $\Delta w_j = \eta \, (R - 1)\Ejr$ (LR$0$, $\mathcal{F}=216.2$, \Fref{fig:results-reinforcement-learning-fitness}C,D).
This rule uses a fixed base line, the maximal reward ($R_\text{max}=1$), rather than the expected reward.
Some runs discover a more sophisticated variant of this rule with a term that decreases the effective learning rate for negative rewards as the agent improves, i.e., when the expected reward increases: $\Delta w_j = \eta \, (1 + R\Ravg)(R - 1)\Ejr$ (LR$1$, $\mathcal{F}=234.2$, \Fref{fig:results-reinforcement-learning-fitness}C,D). % original form: \frac{1}{R}(R + \Ravg)(R - 1)\Ejr
Using this effective learning-rate, this rule achieve higher fitness than the vanilla formulation at the expense of requiring the agent to keep track of the expected reward.

Using the expected reward as a baseline, e.g., $\Delta w_j = \eta \, (R - \Ravg)\Ejr$, is unlikely to yield high-performing solutions: an agent may get stuck in weight configurations in which it continuously receives negative rewards, yet, as it is expecting negative rewards, does not significantly change its weights.
This intuition is supported by our observation that none of the high-performing plasticity rules discovered by our evolutionary search make use of such a baseline, in contrast to previous studies \citep[e.g.,][]{fremaux2016neuromodulated}.
Subtracting the maximal reward, in contrast, can be interpreted as an optimistic baseline \cite[cf.~][Ch2.5]{sutton2018reinforcement}, which biases learning to move away from weight configurations that result in negative rewards, while maintaining weight configurations that lead to positive rewards.
However, a detrimental effect of such an optimistic baseline is that learning is sparse, as it only occurs upon receiving negative rewards, an assumption at odds with behavioral evidence.

In contrast, evolutionary runs with access to separate estimates of the negative and positive rewards discover plasticity rules with efficient baselines, resulting in increased fitness (see Appendix \Fref{sec:app-reward-learning} for the full data).
In the following we discuss four such high-performing plasticity rules with at least $10\%$ higher fitness than LR$0$ (\Fref{fig:results-reinforcement-learning-fitness}D).
We first consider the rule (LR$2$, $\mathcal{F}=242.0$, \Fref{fig:results-reinforcement-learning-fitness}D)
\begin{align}
  % 4931d81e37; $\Delta w_j = \eta \, \Ejr(R - \Ravgplus + \Ravgminus)$, $\mathcal{F}=242.0$
  \Delta w_j = \eta \, [R - (\Ravgplus - \Ravgminus)]\Ejr = \eta (R - \Ravgabs)\Ejr \; ,
  \label{eq:results-reward-learning-lr2}
\end{align}
where we introduced the expected absolute reward $\Ravgabs := \Ravgplus - \Ravgminus = |\Ravgplus| + |\Ravgminus|$, with $\Ravgabs \in [0, 1]$.
Note the difference to the expected reward $\Ravg = \Ravgplus + \Ravgminus$.
Since the absolute magnitude of positive and negative rewards is identical in the considered task, $\Ravgabs$ increases in each trial, starting at zero and slowly converging to one.
Instead of keeping track of the expected reward, the agent can thus simply optimistically increase its baseline with each trial.
Behind this lies the, equally optimistic, expectation that the agent improves its performance over trials.
Starting out as $R\Ejr$ and converging to $(R-1)\Ejr$ this rule combines efficient learning from both positive and negative rewards initially, with improved convergence towards successful weight configuration during late learning.
Note that such a strategy may lead to issues with un- or re-learning.
This may be overcome by the agent resetting the expected absolute reward $\Ravgabs$ upon encountering a new task, similar to a ``novelty'' signal.

Furthermore, our algorithm discovers a variation of this rule (LR$3$, $\mathcal{F}=256.0$, \Fref{fig:results-reinforcement-learning-fitness}D), which replaces $\eta$ with $\eta/(1 + \Ravgplus)$ to decrease the magnitude of weight changes in regions of the weight space associated with high performance.
This can improve convergence properties.

We next consider the rule (LR$4$, $\mathcal{F}=247.2$, \Fref{fig:results-reinforcement-learning-fitness}D):
\begin{align}
  % 6f5a460844; $\Delta w_j = \eta \, (R - 1)(R + E_j + 2 \Ravgplus)$, $\mathcal{F}=247.2$
  \Delta w_j = \eta \left[ \, (R - 1)\Ejr + (R - 1)(R + 2 \Ravgplus) \right] \; .
  \label{eq:results-reward-learning-lr4}
\end{align}
This rule has the familiar form of LR$0$ and LR$1$, with an additional homeostatic term.
Due to the prefactors $R-1$, this rule only changes weights on trials with negative reward.
Initially, the expected reward $\Ravgplus$ is close to zero and the homeostatic term results in potentiation of all synapses, causing more and more neurons to spike.
In particular if initial weights are chosen poorly, this can make learning more robust.
As the agent improves and the expected positive rewards increases, the homeostatic term becomes negative.
In this regime it can be interpreted as pruning all weights until only those are left that do not lead to negative rewards.
This term can hence be interpreted as an adapting action baseline \citep{sutton2018reinforcement}.

Finally, we consider the rule (LR$5$, $\mathcal{F}=254.8$, \Fref{fig:results-reinforcement-learning-fitness}D):
\begin{align}
  % b7ab3994d0, $\Delta w_j = \eta \, (-R\Ravgminus + 2\Ejr)(R\Ravgminus + R - \Ravgplus)$, $\mathcal{F}=254.8$
  \Delta w_j = \eta \left\{ \, 2[R - (\Ravgplus - R \Ravgminus)] \Ejr - [R - (\Ravgplus - R \Ravgminus)] R \Ravgminus \right\} \; .
  \label{eq:results-reward-learning-lr5}
\end{align}
To analyze this seemingly complex rule, we consider the expression for trials with positive and trials with negative reward separately:
\begin{align*}
  R=1: \;\, \Delta w_j^+ =& \, \eta \left\{ \, 2 (1 - \Ravgabs) \Ejr - (1 - \Ravgabs) \Ravgminus \right\} \; , \\
  R=-1: \;\, \Delta w_j^- =& \, \eta \left\{ \, 2(-1 - \Ravg) \Ejr - (1 + \Ravg) \Ravgminus \right\} \; .
\end{align*}
Both expressions contain a ``causal'' term depending on pre- and postsynaptic activity via the eligibility trace, as well as, and a ``homeostatic'' term.
Aside from the constant scaling factor, the causal term of $\Delta w_j^+$ is identical to LR$2$ (\Fref{eq:results-reward-learning-lr2}), i.e., it only causes weight changes early during learning, and converges to zero for later times.
Similarly, the causal term of $\Delta w_j^-$ is initially identical to that of LR$2$ (\Fref{eq:results-reward-learning-lr2}), decreasing weights for connections contributing to wrong decisions.
However it increases in magnitude as the agent improves and the expected reward increases.
The homeostatic term of $\Delta w_j^+$ is potentiating, similarly to LR$4$ (\Fref{eq:results-reward-learning-lr4}): it encourages spiking by increasing all synaptic weights during early learning and decreases over time.
The homeostatic term for negative rewards is also potentiating, but does not vanish for long times unless the agent is performing perfectly ($\Ravgminus \to 0$).
Over time this plasticity rule hence reacts less and less to positive rewards, while increasing weight changes for negative rewards.
The reward-modulated potentiating homeostatic mechanisms can prevent synaptic weights from vanishing and thus encourage exploration for challenging scenarios in which the agent mainly receives negative rewards.

In conclusion, by making use of the separately estimated expected negative and positive rewards in precise combinations with the eligibility trace and the instantaneous reward, our evolving-to-learn approach discovered a variety of reward-based plasticity rules, many of then outperforming previously known solutions \citep[e.g.,][]{urbanczik2009reinforcement}.
The evolution of closed-form expressions allowed us to analyze the computational principles that allow these newly discovered rules to achieve high fitness.
This analysis suggests new mechanisms for efficient learning, for example from ``novelty'' and via reward-modulated homeostatic mechanisms.
Each of these new hypotheses for reward-driven plasticity rules makes specific predictions about behavioral and neuronal signatures that potentially allow us to distinguish between them.
For example LR$2$, LR$3$ and LR$5$ suggest that agents initially learn both from positive and negative rewards, while later they mainly learn from negative rewards.
In particular the initial learning from positive rewards distinguishes these hypotheses from LR$0$, LR$1$, and LR$4$, and previous work \citep{urbanczik2009reinforcement}.
As LR$2$ does not make use of the, separately estimated, expected rewards, it is potentially employed in settings in which such estimates are difficult to obtain.
Furthermore, LR$4$ and LR$5$ suggest that precisely regulated homeostatic mechanisms play a crucial role besides weight changes due to pre- and post-synaptic activity traces.
During early learning, both rules implement potentiating homeostatic mechanisms triggered by negative rewards, likely to explore many possible weight configurations which may support successful behavior.
During late learning, LR$4$ suggests that homeostatic changes become depressing, thus pruning unnecessary or even harmful connections.
In contrast, they remain positive for LR$5$, potentially avoiding catastrophic dissociation between inputs and outputs for challenging tasks.
Besides experimental data from the behavioral and neuronal level, different artificial reward-learning scenarios could further further select for strengths or against weaknesses of the discovered rules.
Furthermore, additional considerations, for example achieving small variance in weight updates \citep{williams1986reinforcement,dayan1991reinforcement}, may lead to particular rules being favored over others.
We thus believe that our new insights into reinforcement learning are merely a forerunner of future experimental and theoretical work enabled by our approach.

\subsection{Evolving an efficient error-driven plasticity rule}
\label{sec:evolving-optimal-error-driven-plasticity}

We next consider a supervised learning task in which a neuron receives information about how far its output is from the desired behavior, instead of just a scalar reward signal as in the previous task.
The widespread success of this approach in machine learning highlights the efficacy of learning from errors compared to correlation- or reward-driven learning \citep{goodfellow2016deep}.
It has therefore often been hypothesized that evolution has installed similar capabilities in biological nervous systems \cite[see, e.g.,][]{marblestone2016toward,whittington2019theories}.

\citet{urbanczik2014learning} introduced an efficient, flexible and biophysically plausible implementation of error-driven learning via multi-compartment neurons.
For simplicity, we consider an equivalent formulation of this learning principle in terms of two point neurons modeled as leaky integrator neurons with exponential postsynaptic currents and stochastic spike generation.
One neuron mimics the somatic compartment and provides a teaching signal to the other neuron acting as the dendritic compartment.
Here, the difference between the teacher and student membrane potentials drives learning:
\begin{align}
  \label{eq:results-error-driven-us}
  \Delta w_j(t) = \eta \, \left[ v(t) - u(t) \right] \psp_j(t) \; ,
\end{align}
where $v$ is the teacher potential, $u$ the student membrane potential, and $\eta$ a fixed learning rate. $\psp_j(t)=(\kappa * s_j)(t)$ represents the the presynaptic spike train $s_j$ filtered by the synaptic kernel $\kappa$.
\Fref{eq:results-error-driven-us} can be interpreted as stochastic gradient descent on an appropriate cost function \citep{urbanczik2014learning} and can be extended to support credit assignment in hierarchical neuronal networks \citep{sacramento2018dendritic}.
For simplicity we assume the student has direct access to the teacher's membrane potential, but in principle one could also employ proxies such as firing rates as suggested in \citet{pfister2010synapses,urbanczik2014learning}.

\begin{figure}
  \centering
  % result hash: 91f21d6d01014991285e5264a59aa967
  % script: ../code/us_learning/create_manuscript_figure.py
  \includegraphics[width=1.\textwidth]{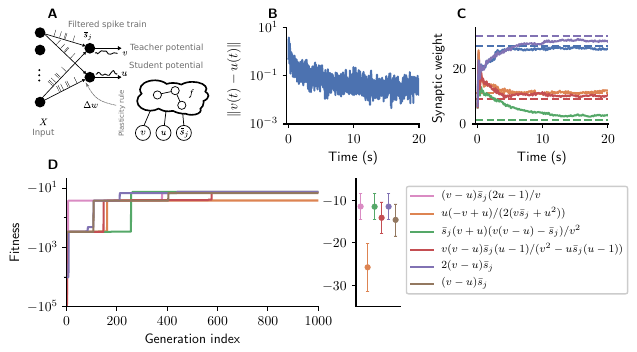}
  \caption{{\bf Cartesian genetic programming evolves efficient error-driven learning rules.}
    {\bf (A)} Network sketch.
    Multiple input neurons with Poisson activity project to two neurons.
    One of the neurons (the teacher) generates a target for the other (the student).
    The membrane potentials of teacher and student as well as the filtered pre-synaptic spike trains are provided to the plasticity rule that determines the weight update.
    {\bf (B)} Root mean squared error between the teacher and student membrane potential over the course of learning using the evolved plasticity rule: $\Delta w_j(t) = \eta \, \left[ v(t) - u(t) \right] \psp_j(t)$.
    {\bf (C)} Synaptic weights over the course of learning corresponding to panel B.
    Horizontal dashed lines represent target weights, i.e., the fixed synaptic weights onto the teacher.
    {\bf (D)} Fitness of the best individual per generation as a function of the generation index for multiple runs of the evolutionary algorithm with different initial conditions.
    Labels represent the rule at the end of the respective run.
    Colored markers represent fitness of each plasticity rule averaged over $15$ validation tasks not used during the evolutionary search; error bars indicate one standard deviation.
  }\label{fig:results-error-driven-learning}
\end{figure}

We consider a one-dimensional regression task in which we feed random Poisson spike trains into the two neurons (\Fref{fig:results-error-driven-learning}A).
The teacher maintains fixed input weights while the student's weights should be adapted over the course of learning such that its membrane potential follows the teacher's (\Fref{fig:results-error-driven-learning}B, C).
The fitness $\mathcal{F}(f)$ of an individual encoding the function $f$ is measured by the root mean-squared error between the teacher and student membrane potential over the simulation duration $T$, excluding the initial $10\%$, averaged over $n_\text{exp}$ experiments:
\begin{align}
  \label{eq:results-error-driven-fitness}
  \mathcal{F}(f) := \frac{1}{n_\text{exp}} \sum_{k=1}^{n_\text{exp}} \sqrt{\int_{0.1 T}^T \text{d}t \left[ v_k(t) - u_k(t) \right]^2} \; .
\end{align}
Each experiment consists of different input spike trains and different teacher weights. The general form of the plasticity rule for this error-driven learning task is given by:
\begin{align}
  \label{eq:results-error-driven-general}
  \Delta w_j = \eta \, f(v, u, \psp_j) \; .
\end{align}
Using CGP with three inputs ($v, u, \psp_j$), we search for plasticity rules that maximize the fitness $\mathcal{F}(f)$.

Starting from low fitness, about half of the evolutionary runs evolve efficient plasticity rules (\Fref{fig:results-error-driven-learning}D) closely matching the performance of the stochastic gradient descent rule of \citet{urbanczik2014learning}.
While two runs evolve exactly \Fref{eq:results-error-driven-us}, other solutions with comparable fitness are discovered, such as
\begin{align}
  \Delta w_j =& \eta (v - u)\psp_j \frac{2u - 1}{v} \; \text{, and}
  \label{eq:results-error-driven-learning-lr0}
  \\
  \Delta w_j =& \eta \psp_j (v + u)\frac{v (v - u) - \psp_j}{v^2} \; .
  \label{eq:results-error-driven-learning-lr1}
\end{align}
At first sight, these rules may appear more complex, but for the considered parameter regime under the assumptions $v \approx u; v, u \gg 1$, we can write them as (see Appendix \ref{sec:app-error-learning}):
\begin{align}
  \Delta w_j = \eta \, c_1 (v - u) \psp_j + c_2 \; ,
\end{align}
with a multiplicative constant $c_1 = \mathcal{O}(1)$ and a negligible additive constant $c_2$.
Elementary manipulations of the expressions found by CGP thus demonstrate the similarity of these superficially different rules to \Fref{eq:results-error-driven-us}.
Consequently, they can be interpreted as approximations of gradient descent.
The constants generally fall into two categories: fine-tuning of the learning rate for the set of task samples encountered during evolution ($c_1$), which could be responsible for the slight increase in performance, and factors that have negligible influence and would most likely be pruned over longer evolutionary timescales ($c_2$).
This pruning could be accelerated, for example, by imposing a penalty on the model complexity in the fitness function, thus preferentially selecting simpler solutions.

In conclusion, the evolutionary search rediscovers variations of a human-designed efficient gradient-descent-based learning rule for the considered error-driven learning task.
Due to the compact, interpretable representation of the plasticity rules we are able to analyze the set of high-performing solutions and thereby identify phenomenologically identical rules despite their superficial differences.

\subsection{Evolving an efficient correlation-driven plasticity rule}
\label{sec:correlation-based-pattern-detection}

\begin{figure}
  % Script: corr_learning_evolution.py
  \includegraphics[width=1.\linewidth]{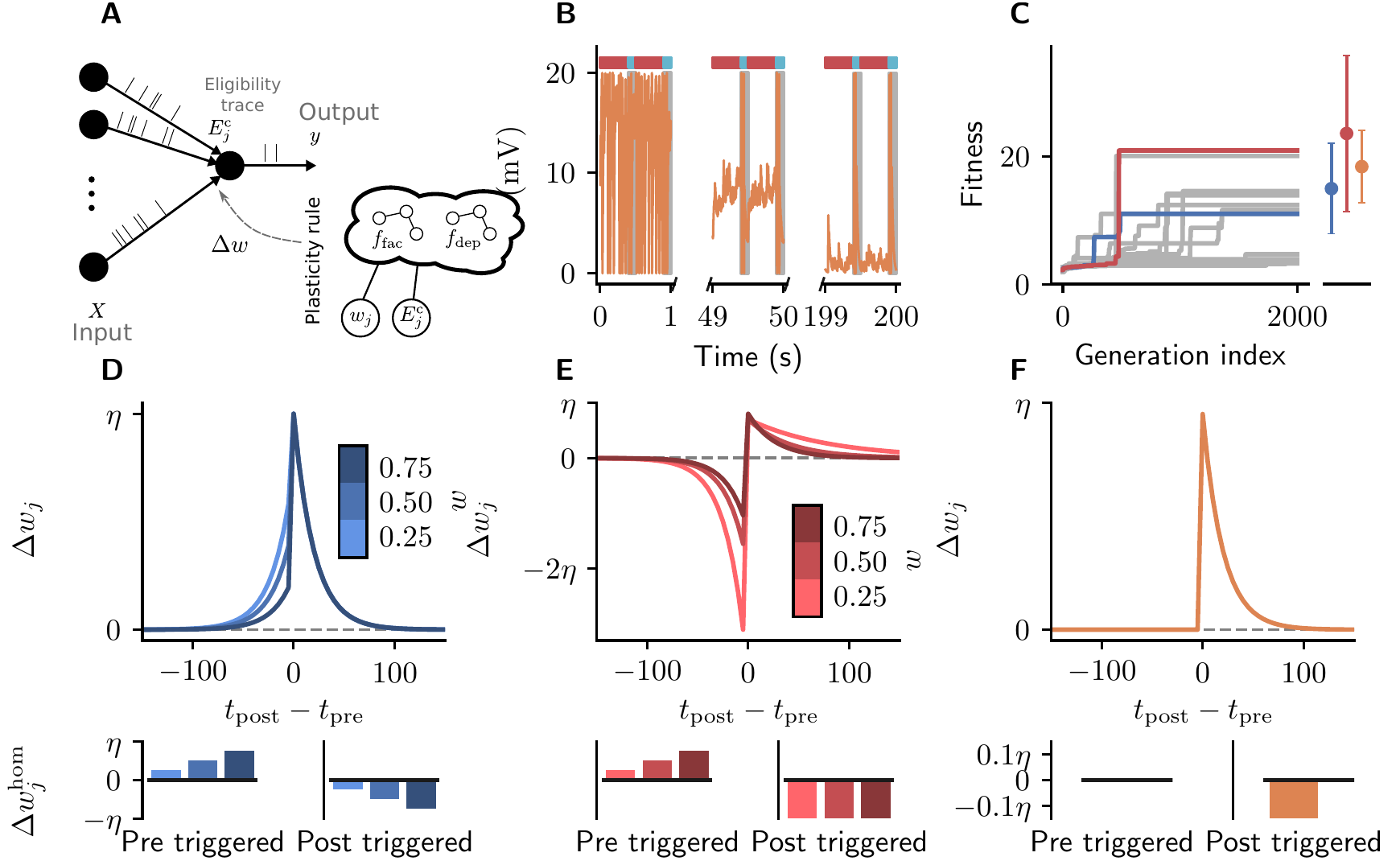}
  \caption{{\bf Cartesian genetic programming evolves diverse correlation-driven learning rules.}
    \textbf{(A)} Network sketch.
    Multiple inputs project to a single output neuron.
    The current synaptic weight $w_j$ and the eligibility trace $E_j^\text{c}$ are provided to the plasticity rule that determines the weight update.
    \textbf{(B)} Membrane potential $u$ of the output neuron over the course of learning using \Fref{eq:stdp_homeostatis}. Gray boxes indicate presentation of the frozen-noise pattern.
    \textbf{(C)} Fitness (\Fref{eq:corr-learning-fitness}) of the best individual per generation as a function of the generation index for multiple runs of the evolutionary algorithm with different initial conditions.
    Blue and red curves correspond to the two representative plasticity rules selected for detailed analysis.
    Blue and red markers represent fitness of the two representative rules and the orange marker the fitness of the homeostatic STDP rule \citep[\Fref{eq:stdp_homeostatis};][]{Masquelier2017}, respectively, on $20$ validation tasks not used during the evolutionary search.
    Error bars indicate one standard deviation over tasks.
    \textbf{(D, E)}: Learning rules evolved by two runs of CGP (\textbf{D}: LR1, \Fref{eq:correlation-lr1}; \textbf{E}: LR2, \Fref{eq:correlation-lr2}).
    \textbf{(F)}: Homeostatic STDP rule \Fref{eq:stdp_homeostatis} suggested by \citep{Masquelier2017}.
    Top panels: STDP kernels $\Delta w_j$ as a function of spike timing differences $\Delta t_j$ for three different weights $w_j$.
    Bottom panels: homeostatic mechanisms for those weights.
    The colors are specific to the respective learning rules (blue for LR1, red for LR2), with different shades representing the different weights $w_j$.
    The learning rate is $\eta=0.01$.
  }\label{fig:results-corr-learning-task}
  % Plot with all learning rules: corr_learning_all_lr.eps
\end{figure}

We now consider a task in which neurons do not receive any feedback from the environment about their performance but instead only have access to correlations between pre- and postsynaptic activity.
Specifically, we consider a scenario in which an output neuron should discover a repeating frozen-noise pattern interrupted by random background spikes using a combination of spike-timing-dependent plasticity and homeostatic mechanisms.
Our experimental setup is briefly described as follows: $N$ inputs project to a single output neuron (\Fref{fig:results-corr-learning-task}A).
The activity of all inputs is determined by a Poisson process with a fixed rate.
A frozen-noise activity pattern of duration $T_\text{pattern}$ is generated once and replayed every $T_\text{inter}\ms$ (\Fref{fig:results-corr-learning-task}B) while inputs are randomly spiking in between.

We define the fitness $\mathcal{F}(f)$ of an individual encoding the function $f$ by the minimal average signal-to-noise ratio ($\SNR$) across $n_\text{exp}$ experiments:
\begin{eqnarray}
  \label{eq:corr-learning-fitness}
  \mathcal{F}(f) &:=& \min_k \left\{ \SNR_k  \,, k\in [1, n_\text{exp}] \right\} \; .
\end{eqnarray}
The signal-to-noise ratio $\SNR_k$, following \citet{Masquelier2017}, is defined as the difference between the maximal free membrane potential during pattern presentation averaged over multiple presentations ($\langle u_{k,i,\text{max}}\rangle$) and the mean of the free membrane potential in between pattern presentations ($\langle u_{k,\text{inter}} \rangle$) divided by its variance ($\mathrm{Var}(v_{k,\mathrm{inter}})$):
\begin{eqnarray}
  \SNR_k &:=& \frac{\langle u_{k,i,\max} \rangle - \langle u_{k,\mathrm{inter}} \rangle}{\mathrm{Var}(u_{k,\mathrm{inter}})} \; .
\end{eqnarray}
The free membrane potential is obtained in a separate simulation with frozen weights by disabling the spiking mechanism for the output neuron.
This removes measurement noise in the signal-to-noise ratio arising from spiking and subsequent membrane-potential reset.
Each experiment consists of different realizations of a frozen-noise pattern and background spiking.

We evolve learning rules of the following general form, which includes a dependence on the current synaptic weight in line with previously suggested STDP rules \citep{gutig2003learning}:
\begin{equation}
  \label{eq:correlation_based_general_lr}
  \Delta w_j^\text{STDP} = \eta  \begin{cases}
    f_{\mathrm{dep}}(w_j, E_j^\text{c}) &\Delta t_j <0 \\
    f_{\mathrm{fac}} (w_j, E_j^\text{c}) &\Delta t_j \geq 0 \; .
  \end{cases}
\end{equation}
Here, $E_j^\text{c}:=\e^{- | \Delta t_j | / \tau}$ represents an eligibility trace that depends on the relative timing of post- and presynaptic spiking ($\Delta t_j= t_{\mathrm{post}} - t_{\mathrm{pre}, j}$) and is represented locally in each synapse \citep[e.g.,][]{morrison2008phenomenological}.
$\eta$ represents a fixed learning rate.
The synaptic weight is bound such that $w_j \in[0, 1]$.
We additionally consider weight-dependent homeostatic mechanisms triggered by pre- and postsynaptic spikes, respectively.
These are implemented by additional functions of the general form:
\begin{equation}
  \label{eq:correlation_based_general_hom}
  \Delta \whom_j = \eta \begin{cases}
    f_\mathrm{pre}^\mathrm{hom}(w_j) & \text{upon presynaptic spike} \\
    f_\mathrm{post}^\mathrm{hom}(w_j) & \text{upon postsynaptic spike} \\
  \end{cases}
\end{equation}
Weight changes are determined jointly by \Fref{eq:correlation_based_general_lr} and \Fref{eq:correlation_based_general_hom} as $\Delta w_j = \Delta w_j^\text{STDP} + \Delta \whom$.
Using CGP, we search for functions $f_\text{dep}$, $f_\text{fac}$, $f_\text{pre}^\text{hom}$, and $f_\text{post}^\text{hom}$ that maximize the fitness $\mathcal{F}(f_\text{dep}, f_\text{fac})$ (\Fref{eq:corr-learning-fitness}).

As a baseline we consider a rule described by \citet{Masquelier2017} (\Fref{fig:results-corr-learning-task}C).
It is a simple additive spike-timing-dependent plasticity (STDP) rule that replaces the depression branch of traditional STDP variants with a postsynaptically-triggered constant homeostatic term $\whom <0$ \citep{Kempter99}.
The synaptic weight of the projection from input $j$ changes according to (\Fref{fig:results-corr-learning-task}G):
\begin{equation}
  \Delta w_j^\text{STDP} = \eta  \begin{cases}
    0 &\Delta t_j <0 \text{ (anticausal interaction)}\\
    E_j^\text{c}  &\Delta t_j \geq 0  \text{ (causal interaction)} \; ,
  \end{cases}
  \label{eq:stdp_homeostatis}
\end{equation}
with homeostatic mechanisms:
\begin{equation}
  \Delta \whom_j = \eta \begin{cases}
    0 & \text{upon presynaptic spike} \\
    \whom & \text{upon postsynaptic spike} \; .
  \end{cases}
  \label{eq:hom_homeostatis}
\end{equation}
To illustrate the result of synaptic plasticity following \Fref{eq:stdp_homeostatis} and \Fref{eq:hom_homeostatis}, we consider the evolution of the membrane potential of an output neuron over the course of learning (\Fref{fig:results-corr-learning-task}C).
While the target neuron spikes randomly at the beginning of learning, its membrane potential finally stays subthreshold in between pattern presentations and crosses the threshold reliably upon pattern presentation.

After $2000$ generations, half of the runs of the evolutionary algorithm discover high-fitness solutions (\Fref{fig:results-corr-learning-task}D).
These plasticity rules lead to synaptic weight configurations which cause the neuron to reliably detect the frozen-noise pattern.
From these well-performing learning rules, we pick two representative examples (\Fref{fig:results-corr-learning-task}D, E) to analyze in detail.
Learning rule $1$ (LR1, \Fref{fig:results-corr-learning-task}D) is defined by the following equations:

\begin{equation}
  \label{eq:correlation-lr1}
  \Delta w_j^\text{STDP} = \eta \begin{cases} -(w_j - 1) E_j^\text{c} &\quad \Delta t_j < 0\\
    E_j^\text{c} &\quad \Delta t_j \geq 0
  \end{cases}\;,
  \quad\quad
  \Delta \whom_j = \eta \begin{cases}
    w_j & \text{upon presyn.~spike} \\
    -w_j & \text{upon postsyn.~spike} \; .
  \end{cases}
\end{equation}
% The simultaneously evolved homeostatic mechanisms take the following form:
% \begin{equation}
%   \label{eq:correlation-lr1-hom}
%   \Delta \whom_j = \eta \begin{cases}
%     w_j & \text{upon presynaptic spike} \\
%     -w_j & \text{upon postsynaptic spike} \; .
%   \end{cases}
% \end{equation}
Learning rule 2 (LR2, \Fref{fig:results-corr-learning-task}E) is defined by the following equations:
\begin{equation}
  \label{eq:correlation-lr2}
  \Delta w_j^\text{STDP} = \eta \begin{cases} -E_j^\text{c}/w_j   &\quad \Delta t_j < 0\\
    (w_j E_j^\text{c})^{w_j} &\quad \Delta t_j \geq 0
  \end{cases}\;,
  \quad\quad
  \Delta \whom_j = \eta \begin{cases}
    w_j & \text{upon presyn.~spike} \\
    -1 & \text{upon postsyn.~spike} \; .
  \end{cases}
\end{equation}
% accompanied by the following homeostatic mechanisms:
% \begin{equation}
%   \label{eq:correlation-lr2-hom}
%   \Delta \whom_j = \eta \begin{cases}
%     w_j & \text{upon presynaptic spike} \\
%     -1 & \text{upon postsynaptic spike} \; .
%   \end{cases}
% \end{equation}
The form of these discovered learning rules and associated homeostatic mechanisms suggests that they use distinct strategies to detect the repeated spatio-temporal pattern.
LR$1$ causes potentiation for small time differences, regardless of whether they are causal or anticausal (note that $-(w_j - 1) \ge 0$ since $w_j \in [0, 1]$).
In the Hebbian spirit, this learning rule favors correlation between presynaptic and postsynaptic firing.
Additionally, it potentiates synaptic weights upon presynaptic spikes, and depresses them for each postsynaptic spike.
In contrast, LR$2$ implements a similar strategy as the learning rule of \citet{Masquelier2017}: it potentiates synapses only for small, positive (causal) time differences.
Additionally, however, it pronouncedly punishes anticausal interactions.
Similarly to LR$1$, its homeostatic component potentiates synaptic weights upon presynaptic spikes, and depresses them for each postsynaptic spike.

Note how both rules reproduce important components of experimentally established STDP traces \citep[e.g.,][]{caporale2008spike}.
Despite their differences both in the form of the STDP kernel as well as the associated homeostatic mechanisms, both rules lead to high fitness, i.e., comparable system-level behavior.

Unlike the classical perception of homeostatic mechanisms as merely maintaining an ideal working point of neurons \citep{davis2001maintaining}, in both discovered plasticity rules these components support the computational goal of detecting the repeated pattern.
By potentiating large weights more strongly than small weights, the pre-synaptically triggered homeostatic mechanisms support the divergence of synaptic weights into strong weights, related to the repeated pattern, and weak ones, providing background input.
This observation suggests that homeostatic mechanisms and STDP work hand in hand to achieve desired functional outcomes.
Experimental approaches hence need to take both factors into account and variations in observed STDP curves should be reconsidered from a point of functional equivalence when paired with data on homeostatic changes.

In conclusion, for the correlation-driven task, the evolutionary search discovered a wide variety of plasticity rules with associated homeostatic mechanisms supporting successful task learning, thus enabling new perspectives for learning in biological substrates.

\section{Discussion}

% what did we do? what did we learn?
Uncovering the mechanisms of learning via synaptic plasticity is a critical step towards understanding brain (dys)function and building truly intelligent, adaptive machines.
We introduce a novel approach to discover biophysically plausible plasticity rules in spiking neuronal networks.
Our meta-learning framework uses genetic programming to search for plasticity rules by optimizing a fitness function specific to the respective task family.
Our evolving-to-learn approach discovers high-performing solutions for various learning paradigms, reward-driven, error-driven, and correlation-driven learning, yielding new insights into biological learning principles.
Moreover, our results from the reward-driven and correlation-driven task families demonstrate that homeostatic terms and their precise interation with plasticity play an important role in shaping network function, highlighting the importance of considering both mechanisms jointly.
This is, to the best of our knowledge, the first demonstration of the power of genetic programming methods in the search for plasticity mechanisms in spiking neuronal networks.

% more complex tasks
The experiments considered here were mainly chosen due to their simplicity and prior knowledge about corresponding plasticity rules that provided us with a high-performance reference for comparison.
Additionally, in each experiment, we restricted ourselves to a constrained set of possible inputs to the plasticity rule.
Here, we chose quantities which have been previously shown to be linked to synaptic plasticity in various learning paradigms, such as reward, low-pass filtered spike trains, and correlations between pre- and postsynaptic activities.
This prior knowledge avoids requiring the evolutionary algorithm to rediscover these quantities but limits the search space, thus potentially excluding other efficient solutions.

% why compact, interpretable expressions?
A key point of E2L is the compact representation of the plasticity rules.
We restrict the complexity of the expressions by three considerations.
First, we assume that effective descriptions of weight changes can be found that are not unique to each individual synapse. This is is a common assumption in computational neuroscience and based on the observation that nature must have found a parsimonious encoding of brain structure, as not every connection in the brain can be specified in the DNA of the organism \citep{zador2019critique}; rather, genes encode general principles by which the neuronal networks and subnetworks are organized and reorganized \citep{risi2010indirectly}.
Our approach aims at discovering such general principles for synaptic plasticity.
Second, physical considerations restrict the information available to the plasticity rule to local quantities, such as pre- and post-synaptic activity traces or specific signals delivered via neuromodulators \citep[e.g.,][]{cox2019striatal,miconi2018backpropamine}.
Third, we limit the maximal size of the expressions to keep the resulting learning rules interpretable and avoid overfitting.

We explicitly want to avoid constructing an opaque system that has high task performance but does not allow us to understand how the network structure is shaped over the course of learning.
Since we obtain analytically tractable expressions for the plasticity rule, we can analyze them with conventional methods, in contrast to approaches representing plasticity rules with ANNs \citep[e.g.,][]{risi2010indirectly,orchard2016evolution,bohnstingl2019neuromorphic}, for which it is challenging to fully understand their macroscopic computation.
This analysis generates intuitive understanding, facilitating communication and human-guided generalization from a set of solutions to different network architectures or task domains.
In the search for plasticity rules suitable for physical implementations in biological systems, these insights are crucial as the identified plasticity mechanisms can serve as building blocks for learning rules that generalize to the actual challenges faced by biological agents.
Rather than merely applying the discovered rules to different learning problems, researchers may use the analytic expressions and prior knowledge to distill general learning principles -- such as the computational role of homeostasis emerging from the present work -- and combine them in new ways to extrapolate beyond the task families considered in the evolutionary search.
Therefore, our evolving-to-learn approach is a new addition to the toolset of the computational neuroscientist in which human intuition is paired with efficient search algorithms.
Moreover, simple expressions highlight the key interactions between the local variables giving rise to plasticity, thus providing hints about the underlying biophysical processes and potentially suggesting new experimental approaches.

% hypothesis testing machine
From a different perspective, while the learning rules found in the experiments described above were all evolved from random expressions, one can also view the presented framework as a hypothesis-testing machine.
Starting from a known plasticity rule, our framework would allow researchers to address questions like: assuming the learning rule would additionally have access to variable $x$, could this be incorporated into the weight updates such that learning would improve?
The automated procedure makes answering such questions much more efficient than a human-guided manual search.
Additionally, the framework is suitable to find robust biophysically plausible approximations for complex learning rules containing quantities that might be non-local, difficult to compute, and/or hard to implement in physical substrates.
In particular, multi-objective optimization is suitable to evolve a known, complex rule into simpler versions while maintaining high task performance.
Similarly, one could search for modifications of general rules that are purposefully tuned to quickly learn within a specific task family, outperforming more general solutions. %from  https://www.youtube.com/watch?v=hJYyTHQaHxY
In each of these cases, prior knowledge about effective learning algorithms provides a starting point from which the evolutionary search can discover powerful extensions.

% exploiting implicit assumptions
The automated search can discover plasticity rules for a given problem that exploit implicit assumptions in the task.
It therefore highlights underconstrained searches, be this due to scarcity of biological data, the simplicity of chosen tasks or the omission of critical features in the task design.
For instance, without asserting equal average spike rates of background and pattern neurons in the correlation-driven task, one could discover plasticity rules that exploit the rate difference rather than the spatio-temporal structure of the input.

% scientific/historic context
Evolved Plastic Artificial Neural Networks \citep[EPANNs; e.g.,][]{soltoggio2018born} and in particular adaptive HyperNEAT \citep{risi2010indirectly}, represent an alternative approach to designing plastic neural networks.
In contrast to our method, however, these approaches include the network architecture itself into the evolutionary search, alongside synaptic plasticity rules.
While this can lead to high-performance solutions due to a synergy between network architecture and plasticity, this interplay has a an important drawback, as in general it is difficult to tease apart the contribution of plasticity from that of network structure to high task performance \citep[cf.][]{gaier2019weight}.
In addition, the distributed, implicit representation of plasticity rules in HyperNEAT can be difficult to interpret, which hinders a deeper understanding of the learning mechanisms.
In machine-learning-oriented applications, this lack of credit assignment is less of an issue.
For research into plasticity rules employed by biological systems, however, it presents a significant obstacle.

% methodological issues/improvements
Future work needs to address a general issue of any optimization method: how can we systematically counter overfitting to reveal general solutions?
A simple approach would increase the number of sample tasks during a single fitness evaluation. However, computational costs increase linearly in the number of samples.
Another technique penalizes the complexity of the resulting expressions, e.g., proportional to the size of the computational graph.
% discuss ``null terms'' as reviewer 1 was unhappy
Besides avoiding overfitting, such a penalty would automatically remove ``null terms'' in the plasticity rules, i.e., trivial subexpressions which have no influence on the expressions' output.
Since it is a priori unclear how this complexity penalty should be weighted against the original fitness measures, one should consider multi-objective optimization algorithms \citep[e.g.,][]{deb2001multi}.

Another issue to be addressed in future work is the choice of the learning rate.
Currently, this value is not part of the optimization process and all tasks assume a fixed learning rate.
The analysis of the reward- and error-driven learning rules revealed that the evolutionary algorithm tried to optimize the learning rate using the variables it had access to, partly generating complex terms that that amount to a variable scaling of the learning rate.
The algorithm may benefit from the inclusion of additional constants which it could, for example, use for an unmitigated, permanent scaling of the learning rate.
However, the dimensionality of the search space scales exponentially in the number of operators and constants, and the feasibility of such an approach needs to be carefully evaluated.
One possibility to mitigate this combinatorial explosion is to combine the evolutionary search with gradient-based optimization methods that can fine-tune constants in the expressions \citep{topchy2001faster,izzo2017differentiable}.

Additionally, future work may involve less preprocessed data as inputs while considering more diverse mathematical operators.
In the correlation-driven task, one could for example provide the raw times of pre- and postsynaptic spiking to the graph instead of the exponential of their difference, leaving more freedom for the evolutionary search to discover creative solutions.
We expect particularly interesting applications of our framework to involve more complex tasks that are challenging for contemporary algorithms, such as life-long learning, which needs to tackle the issue of catastrophic forgetting \citep{french1999catastrophic} or learning in recurrent spiking neuronal networks.
In order to yield insights into information processing in the nervous system, the design of the network architecture should be guided by known anatomical features, while the considered task families should fall within the realm of ecologically relevant problems.

% speed
The evolutionary search for plasticity rules requires a large number of simulations, as each candidate solution needs to be evaluated on a sufficiently large number of samples from the task family to encourage generalization \citep[e.g.,][]{chalmers1991evolution,bengio1992optimization}.
Due to silent mutations in CGP, i.e., modifications of the genotype that do not alter the phenotype, we use caching methods to significantly reduce computational cost as only new solutions need to be evaluated.
However, even employing such methods, the number of required simulations remains large, in the order of $10^3-10^4$ per evolutionary run.
% discussion of computation time and scalability as reviewer 1 was unhappy
For the experiments considered here, the computational costs are rather low, requiring $24-48$ node hours for a few parallel runs of the evolutionary algorithms, easily within reach of a modern workstation.
The total time increases linearly with the duration of a single simulation.
When considering more complex tasks which would require larger networks and hence longer simulations, one possibility to limit computational costs would be to evolve scalable plasticity rules in simplified versions of the tasks and architectures.
Such rules, quickly evolved, may then be applied to individual instances of the original complex tasks, mimicking the idea of ``evolutionary hurdles'' that avoid wasting computational power on low-quality solutions \citep{so2019evolved,real2020automl}.
A proof of concept for such an approach is the delta rule: originally in used in small-scale tasks, it has demonstrated incredible scaling potential in the context of error backpropagation.
Similar observations indeed hold for evolved optimizers \citep{metz2020tasks}.

Neuromorphic systems -- dedicated hardware specifically designed to emulate neuronal networks -- provide an attractive way to speed up the evolutionary search.
To serve as suitable substrates for the approach presented here, these systems should be able to emulate spiking neuronal networks in an accelerated fashion with respect to real time and provide on-chip plasticity with a flexible specification of plasticity mechanisms \citep[e.g.,][]{davies2018loihi,billaudelle2019versatile,mayr2019spinnaker}.

% strong ending
We view the presented methods as a machinery for generating, testing, and extending hypotheses on learning in spiking neuronal networks driven by problem instances and prior knowledge and constrained by experimental evidence.
We believe this approach holds significant promise to accelerate progress towards deep insights into information processing in physical systems, both biological and biologically inspired, with immanent potential for the development of powerful artificial learning machines.

\section{Methods and Materials}

\subsection{Evolutionary algorithm}
\label{sec:methods-evolutionary-algorithm}

We use a $\mu + \lambda$ evolution strategy \citep{beyer2002evolution} to evolve a population of individuals towards high fitness.
In each generation, $\lambda$ new offsprings are created from $\mu$ parents via tournament selection \citep[e.g.,][]{miller1995genetic} and subsequent mutation.
From these $\mu + \lambda$, individuals the best $\mu$ individuals are selected as parents for the next generation (\Fref{alg:evo-alg}).
\begin{algorithm}
  \KwData{Initial random parent Population $P_0 = \{ p_i \} $ of size $\mu$}
  $t \gets 0$ \\
  
  \While{$t < n_{\mathrm{generations}}$}{
    Create new offspring population $Q_{t} = \text{CreateOffspringPopulation}(P_{t})$\\
    Combine parent + offspring populations $R_t = P_t \cup Q_t$\\
    Evaluate fitness of each individual in $R_t$\\
    Pick $P_{t+1} \subset R_t$ best individuals as new parents\\
    $t \gets t + 1$\\
  }
  
  \textbf{Function} $\text{CreateOffspringPopulation}(P)$\\
  \Begin{
    Offspring population $Q = \{\}$ \\
    \While{$|Q| < \lambda$}{
      Choose random subset of $P$ of size $N_{\mathrm{tournament}}$\\
      Choose best individual in the subset and append to $Q$
    }
    \For{$q_i \in Q$}{
      Mutate each gene of $q_i$ with mutation probability $p_\text{mutation}$
    }
    \textbf{Return} $Q$
  }
  
  \caption{Variant of $\mu + \lambda$ evolution strategies used in this study. Note the absence of a crossover step.
  }
  \label{alg:evo-alg}
\end{algorithm}
In this work we use a tournament size of one and a fixed mutation probability $p_\text{mutate}$ for each gene in an offspring individual.
Since in CGP crossover of individuals can lead to significant disruption of the search process due to major changes in the computational graphs \citep{miller1999empirical}, we avoid it here.
In other words, new offspring are only modified by mutations.
We use neutral search \citep{miller2000cartesian}, in which an offspring is preferred over a parent with equal fitness, to allow the accumulation of silent mutations that can jointly lead to an increase in fitness.
As it is computationally infeasible to exhaustively evaluate an individual on all possible tasks from a task family, we evaluate individuals only on a limited number of sample tasks and aggregate the results into a scalar fitness, either by choosing the minimal result or averaging.
We manually select the number of sample tasks to balance computational costs and sampling noise for each task.
In each generation, we use the same initial conditions to allow a meaningful comparison of results across generations.
If an expression is encountered that can not be meaningfully evaluated, such as division by zero, the corresponding individual is assigned a fitness of $-\infty$.

\subsection{HAL-CGP}
\label{sec:methods-python-cgp}

HAL-CGP \citep{schmidt2020hal} (\href{https://github.com/Happy-Algorithms-League/hal-cgp}{https://github.com/Happy-Algorithms-League/hal-cgp}) is an extensible pure Python library implementing Cartesian genetic programming to represent, mutate and evaluate populations of individuals encoding symbolic expressions targeting applications with computationally expensive fitness evaluations.
It supports the translation from a CGP genotype, a two-dimensional Cartesian graph, into the corresponding phenotype, a computational graph implementing a particular mathematical expression.
These computational graphs can be exported as pure Python functions, NumPy-compatible functions \citep{walt2011numpy}, SymPy expressions \citep{meurer2017sympy} or PyTorch modules \citep{paszke2019pytorch}.
Users define the structure of the two-dimensional graph from which the computational graph is generated.
This includes the number of inputs, columns, rows, and outputs, as well as the computational primitives, i.e., mathematical operators and constants, that compose the mathematical expressions.
Due to the modular design of the library, users can easily implement new operators to be used as primitives.
It supports advanced algorithmic features, such as shuffling the genotype of an individual without modifying its phenotype to introduce additional drift over plateus in the search space and hence lead to better exploration \citep{goldman2014analysis}.
The library implements a $\mu + \lambda$ evolution strategy to evolve individuals (see \Fref{sec:methods-evolutionary-algorithm}).
Users need to specify hyperparameters for the evolutionary algorithm, such as the size of parent and offspring populations and the maximal number of generations.
To avoid reevaluating phenotypes that have been previously evaluated, the library provides a mechanism for caching results on disk.
Exploiting the wide availability of multi-core architectures, the library can parallelize the evaluation of all individuals in a single generation via separate processes.

\subsection{NEST simulator}

Spiking neuronal network simulations are based on the 2.16.0 release of the NEST simulator \citep{gewaltig2007nest} (\href{https://github.com/nest/nest-simulator}{https://github.com/nest/nest-simulator}; commit 3c6f0f3).
NEST is an open-source simulator for spiking neuronal networks with a focus on large networks with simple neuron models.
The computationally intensive propagation of network dynamics is implemented in \Cplusplus{} while the network model can be specified using a Python API \citep[PyNEST;][]{eppler2009pynest,zaytsev2014cynest}.
NEST profits from modern multi-core and multi-node systems by combining local parallelization with OpenMP threads and inter-node communication via the Message Passing Interface (MPI) \citep{jordan2018extremely}.
The standard distribution offers a variety of established neuron and plastic synapse models, including variants of spike-timing-dependent plasticity, reward-modulated plasticity and structural plasticity.
New models can be implemented via a domain-specific language \citep{plotnikov2016nestml} or custom \Cplusplus{} code.
For the purpose of this study, we implemented a reward-driven \citep{urbanczik2009reinforcement} and an error-driven learning rule \citep[\Fref{eq:results-error-driven-us};][]{urbanczik2014learning}, as well as a homeostatic STDP rule \citep[\Fref{eq:stdp_homeostatis};][]{Masquelier2017} via custom \Cplusplus{} code.
Due to the specific implementation of spike delivery in NEST, we introduce a constant in the STDP rule that is added at each potentiation call instead of using a separate depression term.
To support arbitrary mathematical expressions in the error-driven (\Fref{eq:results-error-driven-general}) and correlation-driven synapse models (\Fref{eq:correlation_based_general_lr}), we additionally implemented variants in which the weight update can be specified via SymPy compatible strings \citep{meurer2017sympy} that are parsed by SymEngine (\href{https://github.com/symengine/symengine}{https://github.com/symengine/symengine}), a \Cplusplus{} library for symbolic computation.
All custom synapse models and necessary kernel patches are available as NEST modules in the repository accompanying this study (\href{https://github.com/Happy-Algorithms-League/e2l-cgp-snn}{https://github.com/Happy-Algorithms-League/e2l-cgp-snn}).

\subsection{Computing systems}

Experiments were performed on JUWELS (J\"ulich Wizard for European Leadership Science), an HPC system at the J\"ulich Research Centre, J\"ulich, Germany, with $12$ Petaflop peak performance.
The system contains $2271$ general-purpose compute nodes, each equipped with two Intel Xeon Platinum $8168$ processors ($2\text{x}24$cores) and $12\text{x}8$GB main memory.
Compute nodes are connected via an EDR-Infiniband fat-tree network and run CentOS $7$.
Additional experiments were performed on the multicore partition of Piz Daint, an HPC system at the Swiss National Supercomputing Centre, Lugano, Switzerland with $1.731$ Petaflops peak performance.
The system contains $1813$ general-purpose compute nodes, each equipped with two Intel Xeon E5-2695 v4 processors ($2\text{x}18$ cores) and $64$GB main memory.
Compute nodes are connected via Cray Aries routing and communications ASIC with Dragonfly network topology and run Cray Linux Environment (CLE).
Each experiment employed a single compute node.

\subsection{Reward-driven learning task}
\label{sec:methods-reinforcement-learning-task}

We consider a reinforcement learning task for spiking neurons inspired by \citet{urbanczik2009reinforcement}.
Spiking activity of the output neuron is generated by an inhomogeneous Poisson process with instantaneous rate $\phi$ determined by its membrane potential $u$ \citep{pfister2006optimal,urbanczik2009reinforcement}:
\begin{equation}
  \phi(u) := \rho \, e^{\frac{u - u_\text{th}}{\Delta u}} \; .
\end{equation}
Here, $\rho$ is the firing rate at threshold, $u_\text{th}$ the threshold potential, and $\Delta u$ a parameter governing the noise amplitude.
In contrast to \citet{urbanczik2009reinforcement}, we consider an instantaneous reset of the membrane potential after a spike instead of an hyperpolarization kernel.
The output neuron receives spike trains from sources randomly drawn from an input population of size $N$ with randomly initialized weights ($w_\text{initial} \sim \mathcal{N}(0, \sigma_w)$).
Before each pattern presentation, the output neurons membrane potential and synaptic currents are reset.

The eligibility trace in every synapse is updated in continuous time according to the following differential equation \citep{urbanczik2009reinforcement,fremaux2016neuromodulated}:
\begin{align}
  \label{eq:reward-learning-eligibility-trace}
  \tau_\text{M}\dot \Ejr = -\Ejr + \frac{1}{\Delta u} \left[ \sum_{s\in y}\delta(t - s) - \phi(u(t)) \right] \psp_j(t) \; ,
\end{align}
where $\tau_\text{M}$ governs the time scale of the eligibility trace and has a similar role as the decay parameter $\gamma$ in policy-gradient methods \citep{sutton2018reinforcement}, $\Delta u$ is a parameter of the postsynaptic cell governing its noise amplitude, $y$ represents the postsynaptic spike train, and $\psp_j(t) = (\kappa * s_j)(t)$ the presynaptic spike train $s_j$ filtered by the synaptic kernel $\kappa$.
The learning rate $\eta$ was manually tuned to obtain high performance with the suggested by \cite{urbanczik2009reinforcement}.
Expected positive and negative rewards in trial $i$ are separately calculated as moving averages over previous trials \citep{vasilaki2009spike}:
\begin{align}
  \Ravg_{i}^{+/-} = (1 - \frac{1}{m_\text{r}})\Ravg_{i-1}^{+/-} + \frac{1}{m_\text{r}} [R_{i-1}]_\text{+/-} \; ,
\end{align}
where $m_\text{r}$ determines the number of relevant previous trials and $[x]_+ := \text{max}(0, x), [x]_- := \text{min}(0, x)$.
Note that $\Ravgplus \in [0, 1]$ and $\Ravgminus \in [-1, 0]$, since $R \in \{-1, 1\}$. We obtain the average reward as a sum of these separate estimates $\Ravg = \Ravgplus + \Ravgminus; \Ravg \in [-1, 1]$, while the expected absolute reward is determined by their difference $\Ravgabs = \Ravgplus - \Ravgminus; \Ravgabs \in [0, 1]$.

\subsection{Error-driven learning task}
\label{sec:methods-error-driven-learning-task}

We consider an error-driven learning task for spiking neurons inspired by \citet{urbanczik2014learning}.
$N$ Poisson inputs with constant rates ($r_i \sim \mathcal{U}[r_\text{min}, r_\text{max}], i \in [1, N]$) project to a teacher neuron and, with the same connectivity pattern, to a student neuron.
As in \Fref{sec:methods-reinforcement-learning-task}, spiking activity of the output neuron is generated by an inhomogeneous Poisson process.
In contrast to \Fref{sec:methods-reinforcement-learning-task}, the membrane potential is not reset after spike emission.
Fixed synaptic weights from the inputs to the teacher are uniformly sampled from the interval $[w_\text{min}, w_\text{max}]$, while weights to the student are all initialized to a fixed value $w_0$.
In each trial we randomly shift all teacher weights by a global value $w_\text{shift}$ to avoid a bias in the error signal that may arise if the teacher membrane potential is initially always larger or always smaller than the student membrane potential.
Target potentials are read out from the teacher every $\delta t$ and provided instantaneously to the student.
The learning rate $\eta$ was chosen via grid search on a single example task for high performance with \Fref{eq:results-error-driven-us}.
Similar to \citet{urbanczik2014learning}, we low-pass filter weight updates with an exponential kernel with time constant $\tau_\text{I}$ before applying them.

\subsection{Correlation-driven learning task}
\label{sec:methods-correlation-driven-learning-task}

We consider a correlation-driven learning task for spiking neurons similar to \citet{Masquelier2017}: a spiking neuron, modeled as a leaky integrate-and-fire neuron with delta-shaped post-synaptic currents, receives stochastic spike trains from $N$ inputs via plastic synapses.

To construct the input spike trains, we first create a frozen-noise pattern by drawing random spikes $\mathcal{S}_i^{\mathrm{pattern}} \in [0, \Tpattern ],  i \in [0, N-1]$ from a Poisson process with rate $\nu$.
Neurons that fire at least once in this pattern are in the following called ``pattern neurons'', the remaining are called ``background neurons''.
We alternate this frozen-noise pattern with random spike trains of length $\Tinter$ generated by a Poisson process with rate $\nu$ (\Fref{fig:results-corr-learning-task}B).
To balance the average rates of pattern neurons and background neurons, we reduce the spike rate of pattern neurons in between patterns by a factor $\alpha$.
Background neurons have an average rate of $\nu_{\mathrm{inter}} = \nu  \frac{\Tinter}{\Tinter + \Tpattern}$.
We assume that pattern neurons spike only once during the pattern. Thus, they have an average rate of rate of $\nu = \alpha \nu_{\mathrm{inter}}  + \frac{1}{\Tinter + \Tpattern} = \alpha \nu_{\mathrm{inter}} + \nu_{\mathrm{pattern}}$.
Plugging in the previous expression for $\nu_\text{inter}$ and solving for $\alpha$ yields $\alpha = 1 - \frac{\nu_{\mathrm{pattern}}}{\nu_{\mathrm{inter}}}$.
We choose the same learning rate as \citet{Masquelier2017}.
Due to the particular implementation of STDP-like rules in NEST \citep{morrison2007spike}, we do not need to evolve multiple functions describing correlation-induced and homeostatic changes separately, but can evolve only one function for each branch of the STDP window.
Terms in these functions which do not vanish for $E_j^\text{c} \rightarrow 0$ are effectively implementing pre-synaptically triggered (in the acausal branch) and post-synaptically triggered (in the causal branch) homeostatic mechanisms.

\section{Acknowledgments}
We gratefully acknowledge funding from the European Union, under grant agreements 604102, 720270, 785907, 945539 (HBP) and the Manfred St{\"a}rk Foundation.
We further express our gratitude towards the Gauss Centre for Supercomputing e.V.\,(\href{https://www.gauss-centre.eu}{www.gauss-centre.eu}) for co-funding this project by providing computing time through the John von Neumann Institute for Computing (NIC) on the GCS Supercomputer JUWELS at J\"ulich Supercomputing Centre (JSC).
We acknowledge the use of Fenix Infrastructure resources, which are partially funded from the European Union's Horizon 2020 research and innovation programme through the ICEI project under the grant agreement No. 800858.
We would like to thank all participants from the HBP SP9 F\"urberg meetings for stimulating interactions and Tomoki Fukai for initial discussions and support.
We also thank Henrik Mettler and Akos Kungl for helpful comments on the manuscript.
All network simulations carried out with NEST (\href{https://www.nest-simulator.org}{www.nest-simulator.org}).

\section{Competing interests}

The authors declare no competing interests.

\bibliographystyle{apalike}
\bibliography{bibliography}

\newpage

\appendix

\section{Reward-driven learning -- full results}\label{sec:app-reward-learning}

\subsection{Full evolution data for different CGP hyperparameter choices}

\newcommand{\mymark}[1]{\textcolor{red}{\bf #1}}
\setlength{\columnwidthleft}{0.12\textwidth}
\setlength{\columnwidthmiddle}{0.6\textwidth}
\setlength{\columnwidthmiddleconn}{0.18\textwidth}

\begin{tabularx}{\textwidth}{|p{\columnwidthleft}|X|}
  \hline\modelhdr{2}{}{CGP hyperparameter set $0$}\\ \hline
    \textbf{Population} & $\mu=1, p_\text{mutation}=0.035$ \\
  \textbf{Genome} & $n_\text{inputs}=3, n_\text{outputs}=1, n_\text{rows}=1, n_\text{columns}=24, l_\text{max}=24$ \\
  \textbf{Primitives} & Add, Sub, Mul, Div, Const(1.0), Const(0.5) \\
  \textbf{EA} & $\lambda=4, n_\text{breeding}=4, n_\text{tournament}=1, \text{reorder} = \text{true}$ \\
  \textbf{Other} & $\text{max generations}=1000, \text{minimal fitness}=500.0$ \\
  \hline
\end{tabularx}

\noindent
\begin{tabularx}{\textwidth}{|p{\columnwidthleft}|p{\columnwidthleft}|X|}
  \hline\modelhdr{3}{}{Discovered plasticity rules for hyperparameter set $0$}\\ \hline
  \bf Label & \bf Fitness $\mathcal{F}$ & \bf Expression $f$ \\ \hline
  LR$0$ & $216.2$ & $-\Ejr + \Ejr/R$ \\
  LR$1$ & $73.0$ & $(R + {\Ejr}^2)/\Ravg$ \\
  LR$2$ & $216.2$ & $\Ejr(R - 1.0)$ \\
  LR$3$ & $221.6$ & $\Ejr/(2R + (R + 1.0)(R + \Ravg) + 1.0)$ \\
  LR$4$ & $234.2$ & $-\Ejr(R - 1)(R + \Ravg)$ \\
  LR$5$ & $216.2$ & $\Ejr(R - 1)$ \\
  LR$6$ & $69.2$ & $4.0{\Ejr}^2/\Ravg + 2.0\Ejr$ \\
  LR$7$ & $234.2$ & $\Ejr(R - 1)(R + \Ravg)/R$ \\
  \hline
\end{tabularx}

\begin{figure}[h]
  \centering
  \includegraphics[width=.75\textwidth]{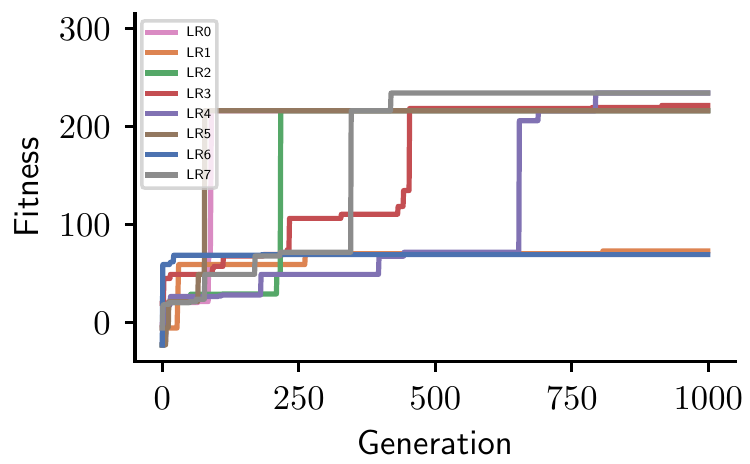}
  \caption{Fitness of best individual per generation as a function of the generation index for multiple runs of the evolutionary algorithm with different initial conditions for hyperparameter set $0$.
  }\label{fig:app-evo-a0dc2d54ba}
\end{figure}

\newpage

\noindent
\begin{tabularx}{\textwidth}{|p{\columnwidthleft}|X|}
  \hline\modelhdr{2}{}{CGP hyperparameter set $1$}\\ \hline
    \textbf{Population} & $\mu=1, p_\text{mutation}=0.035$ \\
  \textbf{Genome} & $n_\text{inputs}=\mymark{4}^*, n_\text{outputs}=1, n_\text{rows}=1, n_\text{columns}=\mymark{12}, l_\text{max}=\mymark{12}$ \\
  \textbf{Primitives} & Add, Sub, Mul, Div, Const(1.0), Const(0.5) \\
  \textbf{EA} & $\lambda=4, n_\text{breeding}=4, n_\text{tournament}=1, \text{reorder} = \text{true}$ \\
  \textbf{Other} & $\text{max generations}=1000, \text{minimal fitness}=500.0$ \\
  \hline
\end{tabularx}
$^*$Red highlights values changed with respect to hyperparameter set $0$. \\

\noindent
\begin{tabularx}{\textwidth}{|p{\columnwidthleft}|p{\columnwidthleft}|X|}
  \hline\modelhdr{3}{}{Discovered plasticity rules for hyperparameter set $1$}\\ \hline
  \bf Label & \bf Fitness $\mathcal{F}$ & \bf Expression $f$ \\ \hline
  LR$0$ & $238.6$ & $(-\Ejr(R + \Ravgminus(R + \Ravgminus)) + \Ejr + \Ravgminus)/(R + \Ravgminus(R + \Ravgminus))$ \\
  LR$1$ & $233.4$ & $\Ejr(R - 1)/(R(R - \Ravgplus))$ \\
  LR$2$ & $217.2$ & $-\Ejr(-R + \Ravgminus + 1.0)$ \\
  LR$3$ & $227.6$ & $R\Ravgminus - \Ejr + \Ejr/R$ \\
  LR$4$ & $247.2$ & $(R - 1.0)(R + \Ejr + 2\Ravgplus)$ \\
  LR$5$ & $198.2$ & $(\Ejr - \Ravgplus - \Ravgminus)/(R + \Ravgplus)$ \\
  LR$6$ & $216.2$ & $\Ejr(R - 1)$ \\
  LR$7$ & $225.8$ & $-\Ejr - \Ravgminus + (R - \Ravgminus)(\Ejr + \Ravgminus)$ \\
  \hline
\end{tabularx}

\begin{figure}[h]
  \centering
  \includegraphics[width=.75\textwidth]{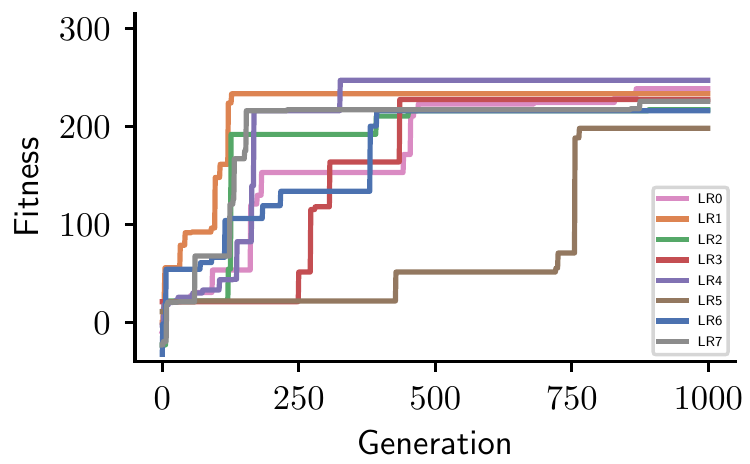}
  \caption{Fitness of best individual per generation as a function of the generation index for multiple runs of the evolutionary algorithm with different initial conditions for hyperparameter set $1$.
  }\label{fig:app-evo-6f5a460844}
\end{figure}

\newpage

\noindent
\begin{tabularx}{\textwidth}{|p{\columnwidthleft}|X|}
  \hline\modelhdr{2}{}{CGP hyperparameter set $2$}\\ \hline
    \textbf{Population} & $\mu=1, p_\text{mutation}=0.035$ \\
  \textbf{Genome} & $n_\text{inputs}=4, n_\text{outputs}=1, n_\text{rows}=1, n_\text{columns}=\mymark{24}^*, l_\text{max}=\mymark{24}$ \\
  \textbf{Primitives} & Add, Sub, Mul, Div, Const(1.0), Const(0.5) \\
  \textbf{EA} & $\lambda=4, n_\text{breeding}=4, n_\text{tournament}=1, \text{reorder} = \text{\textcolor{red}{\bf false}}$ \\
  \textbf{Other} & $\text{max generations}=1000, \text{minimal fitness}=500.0$ \\
  \hline
\end{tabularx}
$^*$Red highlights values changed with respect to hyperparameter set $1$. \\

\noindent
\begin{tabularx}{\textwidth}{|p{\columnwidthleft}|p{\columnwidthleft}|X|}
  \hline\modelhdr{3}{}{Discovered plasticity rules for hyperparameter set $2$}\\ \hline
  \bf Label & \bf Fitness $\mathcal{F}$ & \bf Expression $f$ \\ \hline
  LR$0$ & $127.2$ & $\Ejr/(R + \Ravgplus - \Ravgminus)$ \\
  LR$1$ & $192.0$ & $\Ejr/(R + \Ravgplus)$ \\
  LR$2$ & $216.2$ & $\Ejr(R - 1)$ \\
  LR$3$ & $170.6$ & $(2\Ejr\Ravgminus(R - \Ravgminus) + \Ejr - 1)/(R - \Ravgminus)$ \\
  LR$4$ & $237.6$ & $(-R\Ejr(\Ravgminus + 1) + \Ejr + \Ravgminus)/(R(\Ravgminus + 1))$ \\
  LR$5$ & $233.4$ & $\Ejr(1 - R)/(R - \Ravgplus)$ \\
  LR$6$ & $120.8$ & $(R + \Ravgminus)(\Ejr - \Ravgplus)$ \\
  LR$7$ & $254.8$ & $(-R\Ravgminus + 2\Ejr)(R\Ravgminus + R - \Ravgplus)$ \\
  \hline
\end{tabularx}

\begin{figure}[h]
  \centering
  \includegraphics[width=.75\textwidth]{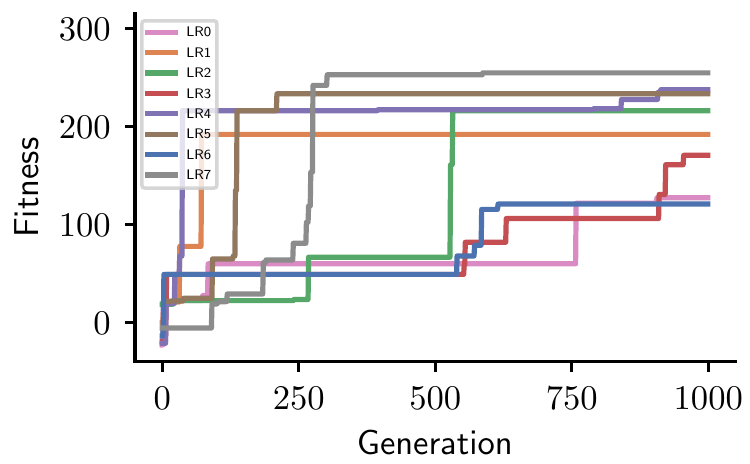}
  \caption{Fitness of best individual per generation as a function of the generation index for multiple runs of the evolutionary algorithm with different initial conditions for hyperparameter set $2$.
  }\label{fig:app-evo-b7ab3994d0}
\end{figure}

\newpage

\noindent
\begin{tabularx}{\textwidth}{|p{\columnwidthleft}|X|}
  \hline\modelhdr{2}{}{CGP hyperparameter set $3$}\\ \hline
    \textbf{Population} & $\mu=1, p_\text{mutation}=0.035$ \\
  \textbf{Genome} & $n_\text{inputs}=4, n_\text{outputs}=1, n_\text{rows}=1, n_\text{columns}=24, l_\text{max}=24$ \\
  \textbf{Primitives} & Add, Sub, Mul, Div, Const(1.0), Const(0.5) \\
  \textbf{EA} & $\lambda=4, n_\text{breeding}=4, n_\text{tournament}=1, \text{reorder} = \text{\textcolor{red}{\bf true}}^*$ \\
  \textbf{Other} & $\text{max generations}=1000, \text{minimal fitness}=500.0$ \\
  \hline
\end{tabularx}
$^*$Red highlights values changed with respect to hyperparameter set $2$. \\

\noindent
\begin{tabularx}{\textwidth}{|p{\columnwidthleft}|p{\columnwidthleft}|X|}
  \hline\modelhdr{3}{}{Discovered plasticity rules for hyperparameter set $3$}\\ \hline
  \bf Label & \bf Fitness $\mathcal{F}$ & \bf Expression $f$ \\ \hline
  LR$0$ & $236.0$ & $\Ejr(-R^3(\Ravgminus + 1) + 1)/R$ \\
  LR$1$ & $242.0$ & $\Ejr(R - \Ravgplus + \Ravgminus)$ \\
  LR$2$ & $242.0$ & $\Ejr(R - \Ravgplus + \Ravgminus)$ \\
  LR$3$ & $227.6$ & $R(\Ejr + \Ravgminus) - \Ejr$ \\
  LR$4$ & $256.0$ & $\Ejr(R - \Ravgplus + \Ravgminus)/(\Ravgplus + 1.0)$ \\
  LR$5$ & $71.0$ & $(\Ravgplus(-R + \Ejr + \Ravgminus(R + \Ravgminus) + \Ravgminus) - \Ravgminus)/\Ravgplus$ \\
  LR$6$ & $216.2$ & $\Ejr(R - 1.0)$ \\
  LR$7$ & $227.8$ & $(\Ejr - \Ravgminus\,^2)(R + \Ravgminus\,^2 - 1.0)$ \\
  \hline
\end{tabularx}

\begin{figure}[h]
  \centering
  \includegraphics[width=.75\textwidth]{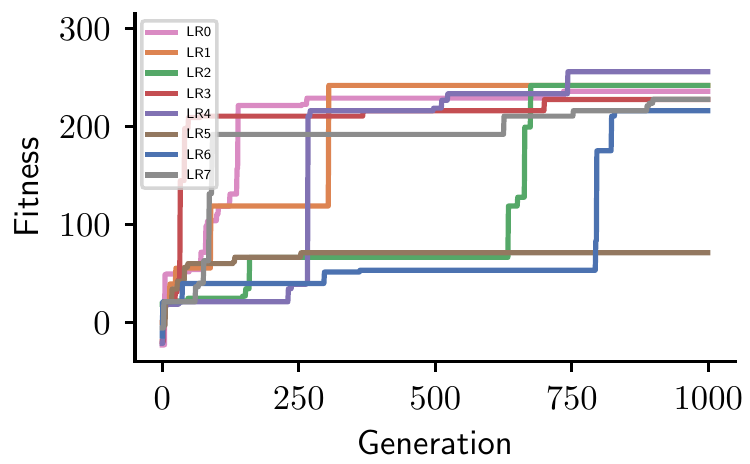}
  \caption{Fitness of best individual per generation as a function of the generation index for multiple runs of the evolutionary algorithm with different initial conditions for hyperparameter set $3$.
  }\label{fig:app-evo-4931d81e37}
\end{figure}

\section{Error-driven learning -- simplification of the discovered rules}\label{sec:app-error-learning}

As in the main manuscript $v$ is the teacher potential, $u$ the student membrane potential, and $\eta$ a fixed learning rate.
$\psp_j(t)=(\kappa * s_j)(t)$ represents the the presynaptic spike train $s_j$ filtered by the synaptic kernel $\kappa$.

We first consider \fref{eq:results-error-driven-learning-lr0}:
\begin{align*}
  \Delta w_j =& \eta (v - u)\psp_j \frac{2u - 1}{v} \\
  =& \eta (v - u)\psp_j \frac{2(v - \delta) - 1}{v} \\
  =& \eta (v - u)\psp_j \left( 2 - 2\underbrace{\frac{\delta}{v}}_{\ll 1} - \underbrace{\frac{1}{v}}_{\approx 0} \right) \\
  \approx& 2 (v - u)\psp_j \; ,
\end{align*}
where we introduced $\delta := v - u$.
From the third to the fourth line we assumed that the mismatch between student and teacher potential is much smaller than their absolute magnitude and that their absolute magnitude is much larger than one.
For our parameter choices and initial conditions this is a reasonable assumption.

We next consider \fref{eq:results-error-driven-learning-lr1}:
\begin{align*}
  \Delta w_j =& \eta \psp_j (v + u)\frac{v (v - u) - \psp_j}{v^2} \\
  =& \eta \psp_j (2v - \delta) \left( \frac{v - u}{v} - \frac{\psp_j}{v^2} \right) \\
  =& \eta \psp_j (2 - \frac{\delta}{v}) \left( (v - u) - \frac{\psp_j}{v} \right) \\
  =& \eta \psp_j \left( \left( 2 - \underbrace{\frac{\delta}{v}}_{\ll 1} \right)(v - u) - 2\underbrace{\frac{\psp_j}{v}}_{\ll 1} + \underbrace{\frac{\delta}{v}}_{\ll 1} \underbrace{\frac{\psp_j}{v}}_{\ll 1} \right) \\
  \approx& 2 (v - u)\psp_j
\end{align*}
As previously, from the third to fourth line we assumed that the mismatch between student and teacher potential is much smaller than their absolute magnitude and that their absolute magnitude is much larger than one.
This implies $\frac{\psp_j}{v} \ll 1$ as $\psp_j \approx \mathcal{O}(1)$ for small input rates.

The additional terms in both \fref{eq:results-error-driven-learning-lr0} and \fref{eq:results-error-driven-learning-lr1} hence reduce to a simple scaling of the learning rate and thus perform similarly to the purple rule in \fref{fig:results-error-driven-learning}.
\section{Correlation-driven learning -- detailed experimental results}\label{sec:app-correlation-learning}

\begin{figure}
  % Script: corr_learning_sim.py
  \includegraphics[width=0.7\textwidth,angle=90,trim=0cm 0cm 0cm 2cm]{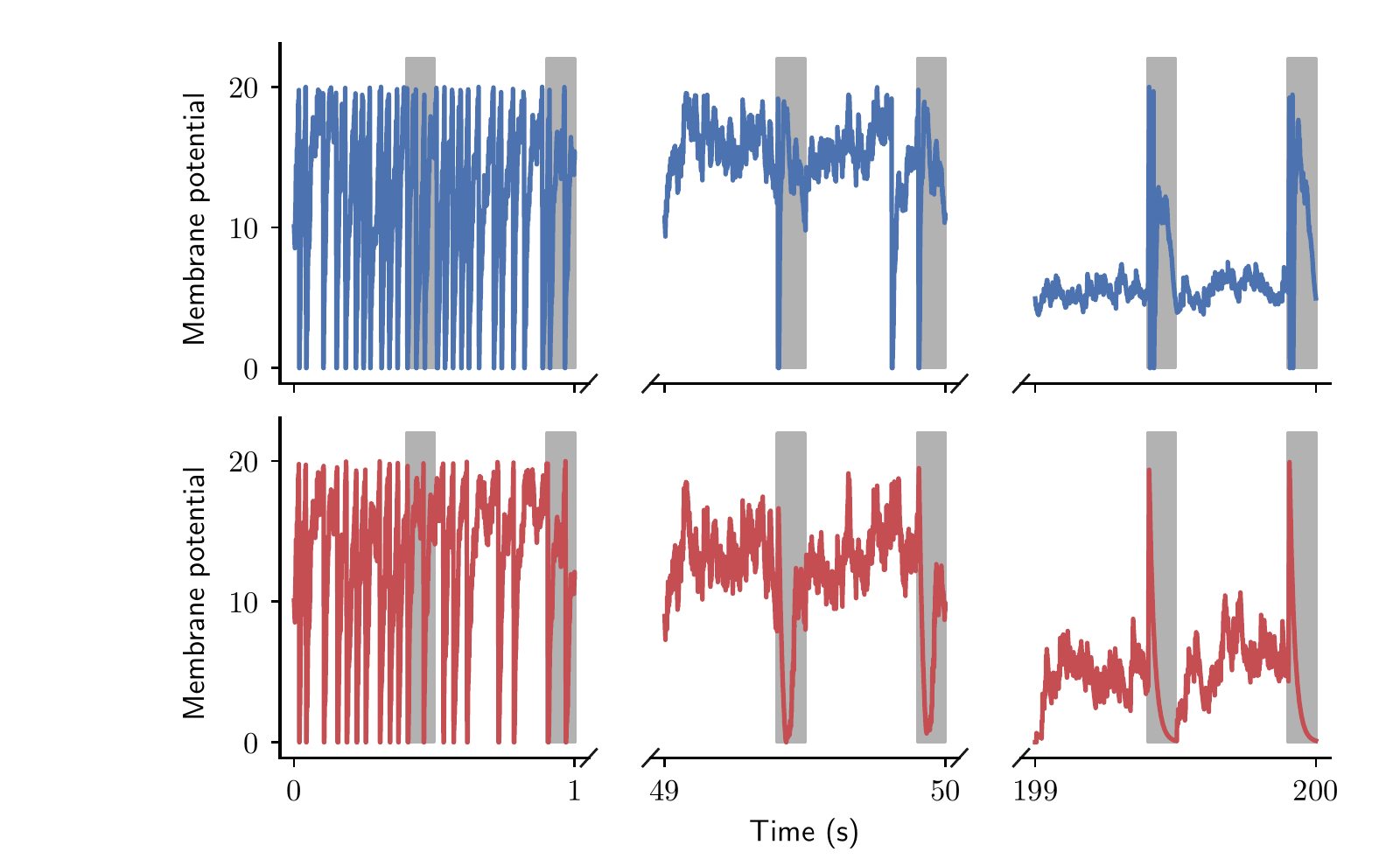}
  \caption{{\bf Evolution of membrane potential for two evolved learning rules.}
    Membrane potential $u$ of the output neuron over the course of learning using the two evolved learning rules LR$1$ (top row, \Fref{eq:correlation-lr1}) and LR$2$ (bottom row, \Fref{eq:correlation-lr2}) (compare \fref{fig:results-corr-learning-task}B).
    Gray boxes indicate presentation of the frozen-noise pattern.
  }
  \label{fig:app-results-corr-learning-task}
  % Plot with all learning rules: corr_learning_all_lr.eps
\end{figure}

\section{Simulation details}
\label{sec:sim-details}
\Fref{tab:supp-nordlie-reward-learning}, \Fref{tab:supp-nordlie-error-learning}, and \Fref{tab:supp-nordlie-corr-learning} summarize the network models used in the experiments \citep[according to][]{nordlie2009towards}.

\begin{table*}

\setlength{\columnwidthleft}{0.2\textwidth}
\setlength{\columnwidthmiddle}{0.6\textwidth}
\setlength{\columnwidthmiddleconn}{0.18\textwidth}

\begin{tabularx}{\textwidth}{|p{\columnwidthleft}|X|}
  \hline\modelhdr{2}{A}{Model summary}\\ \hline
  \textbf{Populations}    & 2 \\ 
  \textbf{Topology}       & \textemdash \\ 
  \textbf{Connectivity}   & Feedforward with fixed connection probability \\ 
  \textbf{Neuron model}   & Leaky integrate-and-fire (LIF) with exponential post-synaptic currents \\
  \textbf{Plasticity}     & Reward-driven \\
  \textbf{Measurements}   & Spikes \\ \hline
\end{tabularx} \\

\begin{tabularx}{\textwidth}{|p{\columnwidthleft}|p{\columnwidthmiddle}|X|}
  \hline\modelhdr{3}{B}{Populations}\\ \hline
  \bf Name & \bf Elements & \bf Size \\ \hline
  Input & Spike generators with pre-defined spike trains (see \Fref{sec:methods-reinforcement-learning-task}) & $N$ \\
  Output & LIF neuron & 1\\ \hline
\end{tabularx} \\

\begin{tabularx}{\textwidth}{|p{\columnwidthleft}|p{\columnwidthmiddleconn}|X|}
  \hline\modelhdr{3}{C}{Connectivity}\\ \hline
  \bf Source & \bf Target & \bf Pattern \\ \hline
  Input & Output & Fixed pairwise connection probability $p$; synaptic delay $d$; random initial weights from $\mathcal{N}(0, \sigma_w^2)$ \\
  \hline
\end{tabularx} \\

\begin{tabularx}{\textwidth}{|p{\columnwidthleft}|X|}
  \hline\modelhdr{2}{D}{Neuron model}\\ \hline
  \textbf{Type} & LIF neuron with exponential post-synaptic currents\\
 \textbf{Subthreshold dynamics} & $\frac{\mathrm{d} u(t)}{\mathrm{d} t}=-\frac{u(t)-E_{\mathrm{L}}}{\tau_{\mathrm{m}}}+\frac{I_{\mathrm{s}}(t)}{C_{\mathrm{m}}}$\hspace*{0.1cm}
if not refractory \newline $u(t)=u_{\mathrm{r}}$\hspace*{1cm}
else 

$I_{\mathrm{s}}(t)=\sum_{i,k} w_{k}\,e^{-(t-t_{i}^{k})/\tau_{\mathrm{s}}}\Theta(t-t_{i}^{k})$, \;
$k$: neuron index, $i$: spike index\\
\textbf{Spiking} & Stochastic spike generation via inhomogeneous Poisson process with intensity $\phi(u) = \rho\, e^{(u-u_\text{th})/\Delta u}$; reset of $u$ to $u_\text{r}$ after spike emission and refractory period of $\tau_\text{r}$ \tabularnewline
\hline 
\end{tabularx}
\begin{tabularx}{\textwidth}{|p{\columnwidthleft}|X|}
  \hline\modelhdr{2}{E}{Synapse model}\\ \hline
  \textbf{Plasticity} & Reward-driven with episodic update (\Fref{eq:reward-learning-gp-rule}, \Fref{eq:reward-learning-gp-rule-2})\\
  \textbf{Other} & Each synapse stores an eligibility trace (\Fref{eq:reward-learning-eligibility-trace})\\
  \hline
\end{tabularx}

\begin{tabularx}{\textwidth}{|p{\columnwidthleft}|X|}
  \hline\modelhdr{2}{F}{Simulation Parameters}\\ \hline
  \textbf{Populations} & $N=50$ \\
  \textbf{Connectivity} & $p=0.8, \sigma_w=10^3\pA$ \\
  \textbf{Neuron model} & $\rho=0.01\Hz, \Delta u=0.2\mV, E_\text{L}=-70\mV, u_\text{r}=-70\mV, u_\text{th}=-55\mV, \tau_\text{m}=10\ms, C_\text{m}=250\pF, \tau_\text{r}=2\ms, \tau_\text{s}=2\ms$ \\
  \textbf{Synapse model} & $\eta=10, \tau_\text{M}=500\ms, d=1\ms$ \\
  \textbf{Input} & $M=30, r=6\Hz, T=500\ms, n_\text{training}=500, n_\text{exp}=10$ \\
  \textbf{Other} & $h=0.01\ms, R \in \{-1, 1\}, m_\text{r} = 100$ \\
  \hline
\end{tabularx}

\begin{tabularx}{\textwidth}{|p{\columnwidthleft}|X|}
  \hline\modelhdr{2}{G}{CGP Parameters}\\ \hline
  \textbf{Population} & $\mu=1, p_\text{mutation}=0.035$ \\
  \textbf{Genome} & $n_\text{inputs}=\{3,4\}, n_\text{outputs}=1, n_\text{rows}=1, n_\text{columns}=\{12, 24\}, l_\text{max}=\{12, 24\}$ \\
  \textbf{Primitives} & Add, Sub, Mul, Div, Const(1.0), Const(0.5) \\
  \textbf{EA} & $\lambda=4, n_\text{breeding}=4, n_\text{tournament}=1, \text{reorder}=\{\text{true}, \text{false}\}$ \\
  \textbf{Other} & $\text{max generations}=1000, \text{minimal fitness}=500$ \\
  \hline
\end{tabularx}

\caption{
  Description of the network model used in the reward-driven learning task (\Fref{sec:methods-reinforcement-learning-task}).
  \label{tab:supp-nordlie-reward-learning}
}

\end{table*}

\begin{table*}

\setlength{\columnwidthleft}{0.2\textwidth}
\setlength{\columnwidthmiddle}{0.6\textwidth}
\setlength{\columnwidthmiddleconn}{0.18\textwidth}

\begin{tabularx}{\textwidth}{|p{\columnwidthleft}|X|}
  \hline\modelhdr{2}{A}{Model summary}\\ \hline
  \textbf{Populations}    & 3 \\ 
  \textbf{Topology}       & \textemdash \\ 
  \textbf{Connectivity}   & Feedforward with all-to-all connections \\ 
  \textbf{Neuron model}   & Leaky integrate-and-fire (LIF) with exponential post-synaptic currents \\
  \textbf{Plasticity} & Error-driven \\
  \textbf{Measurements}   & Spikes, membrane potentials \\ \hline
\end{tabularx} \\

\begin{tabularx}{\textwidth}{|p{\columnwidthleft}|p{\columnwidthmiddle}|X|}
  \hline\modelhdr{3}{B}{Populations}\\ \hline
  \bf Name & \bf Elements & \bf Size \\ \hline
  Input & Spike generators with pre-defined spike trains (see \Fref{sec:methods-error-driven-learning-task}) & $N$ \\
  Teacher & LIF neuron & 1\\
  Student & LIF neuron & 1\\
  \hline
\end{tabularx} \\

\begin{tabularx}{\textwidth}{|p{\columnwidthleft}|p{\columnwidthmiddleconn}|X|}
  \hline\modelhdr{3}{C}{Connectivity}\\ \hline
  \bf Source & \bf Target & \bf Pattern \\ \hline
  Input & Teacher & All-to-all; synaptic delay $d$; random weights $w \sim \mathcal{U}[w_\text{min}, w_\text{max}]$; weights randomly shifted by $w_\text{shift}$ on each trial \\
  Input & Student & All-to-all; synaptic delay $d$; fixed initial weights $w_0$ \\
  \hline
\end{tabularx} \\

\begin{tabularx}{\textwidth}{|p{\columnwidthleft}|X|}
  \hline\modelhdr{2}{D}{Neuron model}\\ \hline
  \textbf{Type} &  LIF neuron with exponential post-synaptic currents\\
 \textbf{Subthreshold dynamics} & $\frac{\mathrm{d} u(t)}{\mathrm{d} t}=-\frac{u(t)-E_{\mathrm{L}}}{\tau_{\mathrm{m}}}+\frac{I_{\mathrm{s}}(t)}{C_{\mathrm{m}}}$

$I_{\mathrm{s}}(t)=\sum_{i,k}J_{k}\,e^{-(t-t_{i}^{k})/\tau_{\mathrm{s}}}\Theta(t-t_{i}^{k})$
$k$: neuron index, $i$: spike index\\
\textbf{Spiking} & Stochastic spike generation via inhomogeneous Poisson process with intensity $\phi(u) = \rho\, e^{(u-u_\text{th})/\Delta u}$; no reset after spike emission \\
\hline 
\end{tabularx}
\begin{tabularx}{\textwidth}{|p{\columnwidthleft}|X|}
  \hline\modelhdr{2}{E}{Synapse model}\\ \hline
  \textbf{Plasticity} & Error-driven with continuous update (\Fref{eq:results-error-driven-us}, \Fref{eq:results-error-driven-general})\\
  \hline
\end{tabularx}

\begin{tabularx}{\textwidth}{|p{\columnwidthleft}|X|}
  \hline\modelhdr{2}{F}{Simulation Parameters}\\ \hline
  \textbf{Populations} & $N=5$ \\
  \textbf{Connectivity} & $w_\text{min}=-20, w_\text{max}=20, w_\text{shift} \sim \{-15,15\}, w_0=5$ \\
  \textbf{Neuron model} & $\rho=0.2\Hz, \Delta u=1.0\mV, E_\text{L}=-70\mV, u_\text{th}=-55\mV, \tau_\text{m}=10\ms, C_\text{m}=250\pF, \tau_\text{s}=2\ms$ \\
  \textbf{Synapse model} & $\eta=1.7, d=1\ms, \tau_\text{I} = 100.0\ms$ \\
  \textbf{Input} & $r_\text{min}=150\Hz, r_\text{max}=850\Hz, T=10,000\ms, n_\text{exp}=15$ \\
  \textbf{Other} & $h=0.01\ms, \delta t=5\ms$ \\
  \hline
\end{tabularx}

\begin{tabularx}{\textwidth}{|p{\columnwidthleft}|X|}
  \hline\modelhdr{2}{G}{CGP Parameters}\\ \hline
  \textbf{Population} & $\mu=4, p_\text{mutation}=0.045$ \\
  \textbf{Genome} & $n_\text{inputs}=3, n_\text{outputs}=1, n_\text{rows}=1, n_\text{columns}=12, l_\text{max}=12$ \\
  \textbf{Primitives} & Add, Sub, Mul, Div, Const(1.0) \\
  \textbf{EA} & $\lambda=4, n_\text{breeding}=4, n_\text{tournament}=1$ \\
  \textbf{Other} & $\text{max generations}=1000, \text{minimal fitness}=0.0$ \\
  \hline
\end{tabularx}

\caption{
  Description of the network model used in the error-driven learning task (\Fref{sec:methods-error-driven-learning-task}).
  \label{tab:supp-nordlie-error-learning}
}

\end{table*}

\begin{table*}

\setlength{\columnwidthleft}{0.2\textwidth}
\setlength{\columnwidthmiddle}{0.6\textwidth}
\setlength{\columnwidthmiddleconn}{0.18\textwidth}

\begin{tabularx}{\textwidth}{|p{\columnwidthleft}|X|}
  \hline\modelhdr{2}{A}{Model summary}\\ \hline
  \textbf{Populations}    & 2 \\ 
  \textbf{Topology}       & \textemdash \\ 
  \textbf{Connectivity}   & Feedforward with all-to-all connections \\ 
  \textbf{Neuron model}   & Leaky integrate-and-fire (LIF) with delta-shaped post-synaptic currents \\
  \textbf{Plasticity} & Correlation-driven \\
  \textbf{Measurements}   & Spikes, membrane potentials \\ \hline
\end{tabularx} \\

\begin{tabularx}{\textwidth}{|p{\columnwidthleft}|p{\columnwidthmiddle}|X|}
  \hline\modelhdr{3}{B}{Populations}\\ \hline
  \bf Name & \bf Elements & \bf Size \\ \hline
  Input & Spike generators with pre-defined spike trains (see \Fref{sec:methods-correlation-driven-learning-task}) & $N$ \\
  Output & LIF neuron & 1\\ \hline
\end{tabularx} \\

\begin{tabularx}{\textwidth}{|p{\columnwidthleft}|p{\columnwidthmiddleconn}|X|}
  \hline\modelhdr{3}{C}{Connectivity}\\ \hline
  \bf Source & \bf Target & \bf Pattern \\ \hline
  Input & Output & All-to-all; synaptic delay $d$; initial weights chosen to yield an approximate initial firing rate of $20\,\mathrm{Hz}$ in the output neuron \newline $w = \frac{u_{\mathrm{th}}}{(N \nu - 2 \sqrt{N \nu}) / (10^{-3}\tau_{\mathrm{m}})}$ \\ \hline
\end{tabularx} \\

\begin{tabularx}{\textwidth}{|p{\columnwidthleft}|X|}
  \hline\modelhdr{2}{D}{Neuron model}\\ \hline
  \textbf{Type} &  LIF neuron with delta-shaped post-synaptic currents\\
 \textbf{Subthreshold dynamics} & $\frac{\mathrm{d} u}{\mathrm{d} t}=-\frac{u-E_{\mathrm{L}}}{\tau_{\mathrm{m}}}+\frac{I_{\mathrm{s}}(t)}{C_{\mathrm{m}}}$\hspace*{0.1cm}
if $\mathrm{\left(t>t^{*}+\tau_{\mathrm{r}}\right)}$\newline $u(t)=u_{\mathrm{r}}$\hspace*{1cm}
else 

$I_{\mathrm{s}}(t)=\sum_{i,k} w_{k}\,\delta(t-t_{i}^{k})$
$k$: neuron index, $i$: spike index\\
\textbf{Spiking} & If $u(t-)<\theta\wedge u(t+)\geq\theta$\newline1. set $t^{*}=t$,
2. emit spike with time stamp $t^{*}$\tabularnewline
\hline 
\end{tabularx}
\begin{tabularx}{\textwidth}{|p{\columnwidthleft}|X|}
  \hline\modelhdr{2}{E}{Synapse model}\\ \hline
\textbf{Plasticity}& Correlation-driven (\Fref{eq:stdp_homeostatis}, \Fref{eq:correlation-lr1}, \Fref{eq:correlation-lr2}) \\
  \hline
\end{tabularx}
\begin{tabularx}{\textwidth}{|p{\columnwidthleft}|X|}
  \hline\modelhdr{2}{F}{Simulation parameters}\\ \hline
\textbf{Neuron model} & $u_{\mathrm{r}}= 0.0\,\mV,\, E_{\mathrm{L}}= 0.0\,\mV, u_{\mathrm{th}}= 20.0\,\mV, \tau_{\mathrm{m}}= 18.0\,\mathrm{ms}$\tabularnewline
\textbf{Synapse model} & $d=1.0\mathrm{ms}$, $W_{\mathrm{max}}= 1.0\mV$, $\tau= 20.0\mV$, $\eta=0.01$ \tabularnewline
\textbf{Experiment} & $N= 1000$, $T= 100\,\mathrm{s}$, $T_{\mathrm{pattern}} = 100.0\,\mathrm{ms}$, $T_{\mathrm{inter}}= 400.0\,\mathrm{ms}$, $\nu = 3.0\,\mathrm{Hz}$, $n_{\mathrm{exp}}=8$ \tabularnewline
\end{tabularx}
\begin{tabularx}{\textwidth}{|p{\columnwidthleft}|X|}
  \hline\modelhdr{2}{F}{CGP parameters}\\ \hline
    \textbf{Population} & $\mu=8, p_\text{mutation}=0.05$ \\
  \textbf{Genome} & $n_\text{inputs}=2, n_\text{outputs}=1, n_\text{rows}=1, n_\text{columns}=5, l_\text{max}=5$ \\
  \textbf{Primitives} & Add, Sub, Mul, Div, Pow, Const(1.0) \\
  \textbf{EA} & $\lambda=8, n_\text{breeding}=8, n_\text{tournament}=1$ \\
  \textbf{Other} & $\text{max generations}=2000, \text{minimal fitness}=10.0$ \\
  \hline
\end{tabularx}

\caption{
  Description of the network model used in the correlation-driven learning task (\Fref{sec:methods-correlation-driven-learning-task}).
  \label{tab:supp-nordlie-corr-learning}
}

\end{table*}

\end{document}